\documentclass[twocolumn,showpacs,groupedaddress,assymb,superscriptaddress,amsmath]{revtex4}
\usepackage{graphicx}
\usepackage{amssymb}
\usepackage{amsmath}

\begin{document}

\title{Dynamics and quantumness of excitation energy transfer through a complex
       quantum network}

\author{M. Qin}
\affiliation{School of Physics and Optoelectronic
Technology, Dalian University of Technology, Dalian 116024,
China}

\author{H. Z. Shen}
\affiliation{School of Physics and Optoelectronic
Technology, Dalian University of Technology, Dalian 116024,
China}

\author{X. L. Zhao}
\affiliation{School of Physics and Optoelectronic
Technology, Dalian University of Technology, Dalian 116024,
China}

\author{X. X. Yi\footnote{Corresponding address: yixx@nenu.edu.cn}}
\affiliation{Center for Quantum Sciences and School of Physics,
Northeast Normal University, Changchun 130024, China}
\date{\today}

\begin{abstract}
Understanding the mechanisms of efficient and robust energy transfer
in organic  systems provides us with new insights for the optimal
design of artificial systems. In this paper, we explore the dynamics
of excitation energy transfer (EET) through a complex quantum
network by a toy model  consisting  of  three sites coupled to
environments. We study how the coherent evolution and the
noise-induced decoherence   work together to reach efficient EET and
illustrate the role of the phase factor attached to the coupling
constant in the EET. By comparing the differences between the
Markovian and non-Markovian dynamics, we discuss the effect of
environment and the spatial structure of system on the dynamics and
the efficiency of EET. A intuitive picture is given to show how the
exciton is transferred through the system. Employing the simple
model, we show the robustness of EET efficiency under the influence
of the environment and elucidate the important role of quantum
coherence in EET. We go further to study the quantum feature of the
EET dynamics by {\it quantumness} and show the importance of quantum
coherence from  a new respect. We calculate the energy current  in
the EET  and its quantumness, results for different system
parameters are presented and discussed.
\end{abstract}

\maketitle
%05.60.Gg: Quantum transport
%03.65.Yz: Decoherence; open systems; quantum statistical methods
%03.67.-a: Quantum information
%71.35.-y: Excitons and related phenomena
\section{Introduction}
The processes of energy and information transfer in  quantum
networks play an important role in many areas of physics, chemistry
and  biology. They  have been identified as central to quantum
mechanics since its early days \cite{Perrin193217}. Under realistic
physical conditions, the coupling of quantum system to environment
is usually unavoidable, which leads to the deterioration of
performance for fundamental mechanical processes in systems such as
computing devices and biological organisms. Recent experiments,
however, find  evidence of long-lived quantum coherence in
conditions that are often defined as hot and wet
\cite{Nagy200616,Gilmore2008112} in the Fenna-Matthews-Olson (FMO)
protein complex of the green-sulfur bacterium $Chlorobium tepidum$
and in the reaction center of the purple bacterium $Rhodobacter
sphaeroides $ \cite{Scholes20113,Grondelle2010463,
Sener2007104,Engel2007446,Lee2007316,Collini2010463,Panitchayangkoon2010107}.

Inspired  by these experimental observations, quantum coherence
across multiple chromophoric sites has been suggested to play a
significant role in achieving the remarkable efficient EET in FMO
and other pigment-protein complexes
\cite{Amerongen2000,Caruso2009131, Ishizaki2009106}. Since then,
many works analyzing energy transport in these systems have been
carried out \cite{Plenio200810,
Ai201286,Hoyer201012,Caruso201285,Chin201012,
Cheng200696,Nazir2009103,Calhoun2009113,Rebentrost200911,
Cui201245,Sarovar20106,Ishizaki2009130,
Yang2011132,Ghosh2011134,Prior2010105,Adolphs200691,Cao2009113},
such as by the quantum network model
\cite{Plenio200810,Ai201286,Hoyer201012,Caruso201285,Chin201012,
Cheng200696,Nazir2009103,Calhoun2009113,Rebentrost200911,Cui201245},
by the hierarchic equation  \cite{Sarovar20106,Ishizaki2009130,
Yang2011132,Ghosh2011134}, by the renormalization  group
\cite{Prior2010105,Adolphs200691}, and by semiquantum
theory\cite{Cao2009113}. Theoretical investigations   found that
dephasing  does not always hinder the efficiency of EET compared
with perfectly quantum coherent system, which is contrary to the
conventional intuition\cite{Faisal20089}. The interplay  of the
quantum coherent evolution and dephasing contributes to the highly
effective EET in light harvesting complexes
\cite{Mohseni2008129,Plenio200810}. Understanding the underlying
mechanism of such a process may assist us in designing novel
nanofabricated structures for quantum transport and optimized solar
cells.

Although there are many progresses made in this field, the interplay
of the quantum coherent evolution and the relaxation process,
\textbf{however, has not been studied in the Haken-Strobl model, in
which the energy relaxations are not included. In this paper, we
introduce the relaxation process into the decoherence terms, and
study the dynamics and the transfer efficiency in a complex
network.} We obtain several \textbf{analytical solutions} to
illuminate how the coherent evolution and the noise-induced
decoherence processes, especially the relaxation process, work
together to reach efficient EET.

Besides, the experimental evidence has shown that the space
distribution of the pigments is very important for the EET
\cite{Grondelle2010463,Sener2007104}. Thus it is essential to take
\textbf{ phase factors} in the site-to-site couplings into account.
We show  how the spatial structure of  the system, embodied in the
site-to-site couplings \cite{Ai201286,Yi201367}, affect the
efficiency of the EET. We also take into account the effect of the
quantum nature of the environment, such as \textbf{Markovianity and
non-Markovianity } on the dynamics
\cite{Rebentrost200911,Rebentrost2009131, Fassioli201012,Tan201258}.
Comparing the result of numerical calculation with recent
experimental results, we analyze the effect of surrounding
environment and inner spatial structure on the EET dynamics and
efficiency. We find that the EET efficiency is robust against
various environmental parameters. To shed more light on the quantum
coherence that has been studied in many literatures in terms of
entanglement \cite{Caruso2009131,Sarovar20106,
Liao201082,Briegel0806,Cai2010104,Thorwart2009478,Caruso201081,Olaya200878},
 we apply a recently
developed measure for quantum coherence called {\it quantumness}
\cite{Giraud201012,Zurek198124,Zurek200375} to study the quantumness
of the EET within the  single excitation subspace. By calculating
the energy current of the transfer dynamics and its quantumness, we
show that quantum coherence indeed play an important role in the
energy transfer.

The remainder of the paper is organized as follows.  In Sec. {\rm
II}, we introduce a model to describe the complex quantum network,
in which  a phase factor is added into the coupling constant. This
 phase factor may come from the spatial structure of the light harvesting system. In
Sec. {\rm III}, we introduce  a Markovian master equation including
the relaxation, dephasing and dissipation to describe the network.
Then we apply the fully connected network to  derive an exact
expression for the EET efficiency. The dependence of the EET on the
initial state and the phase factor in the model are discussed in
details. In Sec. {\rm IV}, we focus on the Markovian and
non-Markovian effects in the dynamics. The effects of the phase
factor and  temperature on the dynamics are also discussed. In Sec.
{\rm V}, we calculate the energy current in the system and discuss
the quantum coherent nature of the dynamics. The effect of the phase
factor and temperature on the current and quantumness are explored.
Sec. {\rm VI} is devoted to concluding remarks.
%%%%%%%%%%%%%%%%%%%%%%%%%%%%%%%%%%%%%%%%%%%%%%%%%%%%%%%%%

\section{Model}
\begin{figure}[h]
\centering
\includegraphics[scale=0.60]{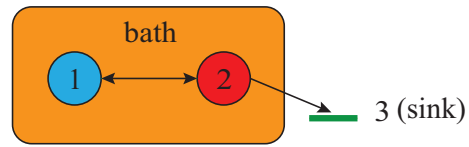}
\caption{(Color online) The schematic representation of three sites
network. The exciton, initially in site $\left| 1 \right\rangle $,
is transferred to site $\left| 2 \right\rangle $ and finally trapped
by the sink $\left| 3 \right\rangle $. } \label{model:}
\end{figure}

We discuss a network of three sites, two of them  $\left| 1
\right\rangle$ and $\left| 2 \right\rangle$   may support
excitations, which can hop from one site to the other (see
Fig.~\ref{model:}). The two sites interact with the environment, and
one of the sites (say site 2) is connected to the third site,  sink
$\left| 3 \right\rangle$. Once excitations fall into the sink, they
can not escape. This network is a model simplified from the
structure of the FMO complex, a network of seven coupled sites, each
of which can be treated as a  two-level system. The total
Hamiltonian in the single exciton manifold includes three parts, $H
= {H_S} + {H_{SB}} + {H_B}$. The system part is given by,
\begin{eqnarray}
{H_S} = {\varepsilon _1}\sigma _1^ + \sigma _1^ -  + {\varepsilon _2}\sigma _2^ + \sigma _2^
-  + J\left( {{e^{ - i\varphi  }}\sigma _1^ + \sigma _2^ -  + {e^{i\varphi  }}\sigma _2^
+ \sigma _1^- } \right),
\label{system H}
\end{eqnarray}
where $\sigma _j^ +$ and $\sigma _j^ -$ are the raising and lowering
operators for site $j$, defined by $\sigma _j^ +  = \left| j
\right\rangle \left\langle 0 \right|$ and $\sigma _j^ -  = \left| 0
\right\rangle \left\langle j \right|$, where $\left| 0
\right\rangle$ denotes the zero exciton state of the system and
$\left| j \right\rangle$ represents  one excitation on  site $j$.
${\varepsilon _j}$ is the on-site energy of site $j$, and ${J_{ij}}$
is designated as the excitonic coupling between site $i$ and $j$,
where ${J_{ij}} = {J_{ji}} = J$ is assumed. A   phase factor ${e^{ -
i\varphi }}$ with $\varphi $ a real number is added to the
inter-site coupling $J$, making the coupling constant complex.

The Hamiltonian of the environment, which modulates  site energy
fluctuation and brings   decoherence, is usually modeled as a set of
quantum  oscillators,
\begin{eqnarray}
{H_B} = \sum\limits_j {\sum\limits_{m = 1,2} {{\omega _{m,j}}b_{m,j}^\dag {b_{m,j}}} },
\label{environment H}
\end{eqnarray}
where $b_{m,j}^\dag$ $({b_{m,j}})$ is the creation  (annihilation)
operator of the $j$-th mode with frequency ${\omega _j}$ in the
bath, we suppose that each site is separately coupled  with its own
environment. Thus the system-environment interection Hamiltonian
reads \cite{Liao201082}
\begin{eqnarray}
{H_{SB}} = \sum\limits_j {\sum\limits_{m = 1,2} {{g_{m,j}}(b_{m,j}^\dag  +
{b_{m,j}})} } \left| m \right\rangle \left\langle m \right|,
\label{interaction H}
\end{eqnarray}
where ${{g_{m,j}}}$ denotes the coupling constant of  the $m$-th
site to its bath mode $j$.

The transport dynamics of a single excitation is described by a
master equation  that includes coherent evolution, relaxation,
dissipation and dephasing. In addition, the exciton can be trapped
in the sink. Once the excitation is trapped, it can not escape.
\section{EET dynamics in Markovian case}
Consider a Markovian environment,  the dynamics of the reduced
density matrix of the system is given by
\begin{eqnarray}
\begin{aligned}
\dot \rho (t) =&  - i[{H_S},\rho ] + {L_\kappa }(\rho (t)) \\
& + {L_\gamma }(\rho (t))+ {L_\Gamma }(\rho (t)) + {L_s}(\rho (t)),
\label{master equation 1}
\end{aligned}
\end{eqnarray}
where the Liouvilian  on the right-hand side in Eq. (\ref{master
equation 1})  describes   coherent dynamics, relaxation, dephasing,
dissipation, and sink,   respectively.

The relaxation   (with rate ${\Gamma _{12}}({\Gamma _{21}})$) is
described  by Lindblad super-operators ${L_\kappa }$
\cite{Rebentrost200911,Tan201258},
\begin{eqnarray}
\begin{aligned}
{L_\kappa }(\rho (t)) =& {\rm{ }} {\Gamma _{12}} [\sigma _2^ + \sigma _1^ - \rho (t)\sigma _1^ + \sigma _2^ - \\
&- \frac{1}{2}\left\{ {\rho (t),\sigma _1^ + \sigma _2^ - \sigma _2^ + \sigma _1^ - } \right\}]\\
&+ {\Gamma _{21}}[\sigma _1^ + \sigma _2^ - \rho (t)\sigma _2^ + \sigma _1^ - \\
&- \frac{1}{2}\left\{ {\rho (t),\sigma _2^ + \sigma _1^ + \sigma _1^
- \sigma _2^ - } \right\}], \label{Lindblad kappa}
\end{aligned}
\end{eqnarray}
where $\left\{ {A,B} \right\}$ denotes an anticommutator, while
${\Gamma _{12}}$ denotes  exciting process from $\left| 1
\right\rangle$ to $\left| 2 \right\rangle$ and ${\Gamma _{21}}$
denotes  decay process from $\left| 2 \right\rangle$ to $\left| 1
\right\rangle$.

The dephasing process (with rate ${{\gamma _j}}$) that destroys the
coherence can be described by,
\begin{eqnarray}
\begin{aligned}
{L_\gamma }(\rho (t)) = \sum\limits_{j = 1,2}  {{\gamma _j}} [\sigma
_j^ + \sigma _j^ - \rho (t)\sigma _j^ + \sigma _j^ - -
\frac{1}{2}\left\{ {\rho (t),\sigma _j^ + \sigma _j^ - } \right\}].
\label{Lindblad gamma 1}
\end{aligned}
\end{eqnarray}
Here, we treat pure dephasing in a way equivalent to the well-known
Haken-Strobl model, in which pure dephasing is considered
phenomenologically in terms of a classical fluctuating field
\cite{Mohseni2008129,Plenio200810}.

Additionally, two processes that lead to irreversible loss of energy
are the dissipation and the sink. These effects are characterized
 by Lindblad operators \cite{Mohseni2008129,Plenio200810}
\begin{eqnarray}
\begin{aligned}
{L_\Gamma }(\rho (t)) = \sum\limits_{j = 1,2}  {{\Gamma _j}[\sigma
_j^ - \rho (t)\sigma _j^ +  - \frac{1}{2}\left\{ {\rho (t),\sigma
_j^ + \sigma _j^ - } \right\}]}
\end{aligned}
\end{eqnarray}
and
\begin{eqnarray}
\begin{aligned}
{L_s}(\rho (t)){\rm{ }} = &{\Gamma _s}[\sigma _3^ + \sigma _2^ - \rho (t)\sigma _2^ + \sigma _3^ -  \\
&- \frac{1}{2}\left\{ {\rho (t),\sigma _2^ + \sigma _3^ - \sigma _3^ + \sigma _2^ - } \right\}],
\label{Lindblad s}
\end{aligned}
\end{eqnarray}
where ${{\Gamma _j}}$ and ${\Gamma _s}$ denote the rates of
dissipation and trapping, respectively. In Ref.
\cite{Rebentrost200911,Tan201258,Mohseni2008129,Olaya200878}, these
two processes are taken into account by adding a non-Hermitian
Hamiltonians ${H_{loss}} = {{ - i} \mathord{\left/
 {\vphantom {{ - i} 2}} \right.
 \kern-\nulldelimiterspace} 2}\sum\limits_{j = 1,2,s}
 {{\Gamma _j}\left| j \right\rangle }
 \left\langle j \right|$ to the system Hamiltonian (\ref{system H}),
which is equivalent to our treatment when the sandwich term can be
ignored.

The efficiency $P$ of EET  is quantified  by the population
transferred to the sink $\left| 3 \right\rangle$ from the site
$\left| 2 \right\rangle$
\begin{eqnarray}
\begin{aligned}
P = \rho {}_{33}(T) = {\Gamma _s}\int_0^T {\rho {}_{22}(t)} dt,
\label{efficiency}
\end{aligned}
\end{eqnarray}
which will be used as a  measure  of the transport efficiency. Thus,
the  question we focus on is as follows: In a given time $T$, what
is the probability transferred from site $\left| 1 \right\rangle $
to the sink $\left| 3 \right\rangle $ and how the spatial structure
of the system and the surrounding environment affect this transfer.

To study how the   coherent   evolution and the noise-induced
decoherence processes  work together  to reach efficient EET and
illustrate the role of the phase factor in the coupling constant, we
adopt a uniform and non-uniform two-level FCN to
help us identify some mechanisms that underlie the energy
transport\cite{Caruso2009131,Chin201012}. A uniform FCN means,
${\varepsilon _1} = {\varepsilon _2} = \varepsilon$, ${\Gamma _1} =
{\Gamma _2} = \Gamma$ and ${\gamma _1} = {\gamma _2} = \gamma$. Given that the exciton is initially in site $\left| 1 \right\rangle$, then
 an exact analytical solution of
$P$ (see the Appendix) for the case of $T = \infty $ can be obtained
\begin{eqnarray}
\begin{aligned}
P = \frac{{{\Gamma _s}}}{{(2\Gamma  + {\Gamma _s}) + \frac{{A\Gamma (\Gamma  + {\Gamma _s}
+ {\Gamma _{21}} - {\Gamma _{12}})}}{{(4{J^2} +
A{\Gamma _{12}})}}}},
\label{solution 1}
\end{aligned}
\end{eqnarray}
where $A = 2\gamma  + 2\Gamma  + {\Gamma _s} + {\Gamma _{12}} +
 {\Gamma _{21}}$. For the case of a \textbf{non-uniform} FCN, ${\varepsilon _1} \ne
{\varepsilon _2}$ while ${{\gamma _j}}$ and ${{\Gamma _j}}$ are the
same on every site. With those notations, the transfer efficiency is
given by,
\begin{eqnarray}
\begin{aligned}
P = \frac{{{\Gamma _s}}}{{(2\Gamma  + {\Gamma _s}) +
\frac{{\Gamma ({A^2} + 16{\Delta ^2})(\Gamma  + {\Gamma _s} + {\Gamma _{21}} -
{\Gamma _{12}})}}{{(4{J^2}A + {\Gamma _{12}}({A^2} +
 16{\Delta ^2}))}}}},
\label{solution 2}
\end{aligned}
\end{eqnarray}
where the on-site energy gap is defined by $\Delta  =
{{({\varepsilon _2} - {\varepsilon _1})} \mathord{\left/
 {\vphantom {{({\varepsilon _2} - {\varepsilon _1})} 2}} \right.
 \kern-\nulldelimiterspace} 2}$. It is notable that, although
 the phase factor $\varphi  $ exists in the system
Hamiltonian of Eq. (\ref{system H}), the final efficiency  $P$ does
not depend on
 $\varphi  $, which is in agreement with the findings in
 Ref. \cite{Ai201286}. This is not the case,
 however, if the exciton is initially in a
 superposition of $\left| 1 \right\rangle $ and $\left| 2 \right\rangle $, as we show below.

\begin{figure*}[htbp]
\centering
\includegraphics[scale=0.355]{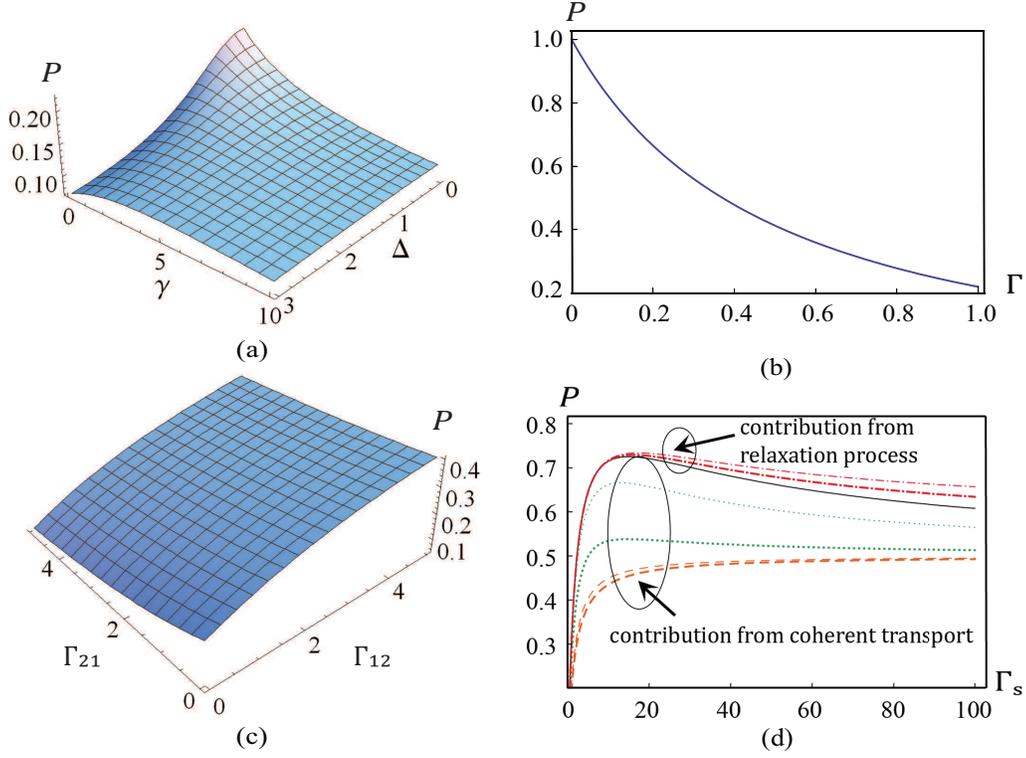}
\caption{(Color online) The transfer efficiency $P$. (a) $P$ vs $\gamma $ and
$\Delta $. (b) $P$ vs $\Gamma $. (c) $P$ vs
${\Gamma _{12}}$ and ${\Gamma _{21}}$. (d) $P$ vs ${\Gamma _s} $. Unless
otherwise noted, the other parameters chosen are $\Gamma = 1$,
$\gamma = 0.1$, $J = 1$, ${\Gamma _s} = 1$, ${\Gamma _{12}} =
{\Gamma _{21}} = 1$, $\Delta  = 0$. In (d), the parameters is as follows: $J = 0$, ${\Gamma _{12}} =
{\Gamma _{21}} = 1$ for thick orange-dashed, $J = 0.5$, ${\Gamma _{12}} =
{\Gamma _{21}} = 1$ for thin orange-dashed, $J = 1.5$, ${\Gamma _{12}} =
{\Gamma _{21}} = 1$ for thick green-dotted, $J = 3$, ${\Gamma _{12}} =
{\Gamma _{21}} = 1$ for thin green-dotted, $J = 4$, ${\Gamma _{12}} =
{\Gamma _{21}} = 1$ for black-solid, $J = 4$, ${\Gamma _{12}} =
{\Gamma _{21}} = 1.2$ for thick red-dot-dashed, ${\Gamma _{12}} =
{\Gamma _{21}} = 1.4$ for thin red-dot-dashed.} \label{parameters:}
\end{figure*}

The dephasing process (represented by the term with ${{\gamma
_j}}$) does not assist the EET as expected,  see Eq. (\ref{solution
1}). It is not the case, however, when we consider a non-uniform
network, see Eq. (\ref{solution 2}), in this case $P$ is not a
monotonic function of $\gamma $. We also observe from
Fig.~\ref{parameters:} (a) that, dephasing facilitates the
transition only when $\Delta $ is  larger than a certain value. The
physics behind this difference is as follows. The range of on-site
energies of $\left| 1 \right\rangle $ and $\left| 2 \right\rangle $
are broadened  owing to the dephasing, which leads to the overlap of
the two sites in energy. If the dephasing rate continues to
increase, the effect of resonant mode decreases as the energy of
each site is distributed over a very large interval. When $\Delta $
is quite small, or even $\Delta  = 0$ (uniform case), resonant mode
is available without the presence of broadened site energy.

From Eq. (\ref{solution 1}) or Eq. (\ref{solution 2}), it is easy to
conclude that the dissipation rate ${{\Gamma _j}}$ simply decreases
the efficiency $P$ (see Fig.~\ref{parameters:} (b)). The dissipation
process leads to irreversible loss of energy via site $j$ to the
environment which obviously  against the increasing of the EET
efficiency.

\textbf{The influence of dephasing and  dissipation in absence of
relaxation (i.e.,${\Gamma _{12}} = {\Gamma _{21}} = 0$) on the
transfer efficiency in a similar network has been explored
extensively in the previous works
\cite{Plenio200810,Ai201286,Hoyer201012,Caruso201285,Rebentrost200911}.}
\textbf{In the following,  we are interested in   new results
obtained when the relaxation process (usually ignored in
Haken-Strobl model)
(\cite{Plenio200810,Ai201286,Hoyer201012,Caruso201285,Rebentrost200911})
is taken into account. We find from Fig.~\ref{parameters:} (c) that
$P$ monotonically increases with ${\Gamma _{12}}$, but decreases
with ${\Gamma _{21}}$,  this finding can also be  obtained from Eq.
(\ref{solution 1}) or Eq. (\ref{solution 2}) by taking  the
derivative of $P$ with respect to ${\Gamma _{12}}$ or ${\Gamma
_{21}}$. The observations can be understood as follows. Relaxation
process represents thermal equilibration of the exciton. The
diffusion of the exciton from $\left| 1\right\rangle$ to $\left| 2
\right\rangle$ obviously increases the population of site $\left| 2
\right\rangle$. If $J = 0$ is set, the efficiency $P =
0$\cite{Plenio200810}, since no process except coherent transfer
(with coupling constant $J$) transfers excitons from $\left| 1
\right\rangle $ to $\left| 2 \right\rangle $. When consider the
relaxation process, the situation changes. For $J = 0$, Eq.
(\ref{solution 1}) becomes
\begin{eqnarray}
\begin{aligned}
P = \frac{{{\Gamma _{12}}{\Gamma _s}}}{{{\Gamma ^2}  + {\Gamma
_{21}}{\Gamma _s} + \Gamma \left( {{\Gamma _{12}} + {\Gamma _{21}} +
{\Gamma _s}} \right)}}. \label{solution 3}
\end{aligned}
\end{eqnarray}
This suggests that  excitons may be transferred  to the sink, even
though coherent transfer characterized by $J$ is zero. As the
excitons can be transferred via relaxation process,  coherent
transport and relaxation process work together to get a highly EET
efficiency.}

\textbf{Fig.~\ref{parameters:} (d) shows the dependence of the
transfer efficiency $P$ on the trapping rate ${{\Gamma _s}}$ and how
these two processes work together to reach efficient EET. In the
dashed, dotted and solid lines, the relaxation rates are the same,
so the improvement in the efficiency  is attributed to the coherent
transfer. The red-dot-dashed and black-solid lines share the same
values of $J$, but the relaxation rates are different. The
improvement of the efficiency is attributed to the relaxation
process. Note that the thick orange-dashed and thin orange-dashed
curves rise monotonically, which  were not predicted in previous
works \cite{Plenio200810,Ai201286,Hoyer201012,Caruso201285}. The
other curves ascend first and descend then. This is due to the
relaxation ignored in the literature. This can be explained by
analyzing Eq. (\ref{solution 1}). Taking derivative of $P$ with
respect to ${\Gamma _s}$ (see Apendix B), we observe that, for a
fixed $\Gamma$, ${\Gamma _{12}}$, ${\Gamma _{21}}$ and $J$, when
${J^2} < {\Gamma _{12}}(\Gamma  + {\Gamma _{12}} + {\Gamma _{21}})$,
$P$ is a monotonic function of ${{\Gamma _s}}$, it takes a maximal
value $P = \frac{{{\Gamma _{12}}}}{{{\Gamma _{12}} + \Gamma}}$ with
${\Gamma _s} \to \infty $, this is shown  by the thick orange-dashed
and thin orange-dashed curves in Fig.~\ref{parameters:} (d). In this
case, the relaxation dominates the transfer. When the trapping rate
${{\Gamma _s}}$ is very small, site $\left| 2 \right\rangle $
couples weakly to the sink $\left| 3 \right\rangle $, excitations
can  rarely reach the sink in this situation, so the efficiency is
very low. When ${{\Gamma _s}}$ is very large, the efficiency can be
improved significantly. The highest efficiency   is determined by
the contribution of incoherence process  given by $P =
\frac{{{\Gamma _{12}}}}{{{\Gamma _{12}} + \Gamma}}$. When ${J^2} >
{\Gamma _{12}}(\Gamma  + {\Gamma _{12}} + {\Gamma _{21}})$, $P$ is
not a monotonic function of ${{\Gamma _s}}$ and it also takes $P =
\frac{{{\Gamma _{12}}}}{{{\Gamma _{12}} + \Gamma}}$ with  ${\Gamma
_s} \to \infty $, but there exists an optimized value of ${{\Gamma
_s}}$ (Fig.~\ref{parameters:} (d)), leading to a efficiency  larger
than $\frac{{{\Gamma _{12}}}}{{{\Gamma _{12}} + \Gamma}}$. From
Fig.~\ref{parameters:} (d), the improvement of efficiency comes from
coherent transfer. In this case, the coherent transfer dominates the
transfer. When ${{\Gamma _s}}$ is very small, the efficiency is also
very low. When ${{\Gamma _s}}$ increases, $P$ increases gradually to
a maximum where the contribution from coherent transfer is brought
into full play and excitation transferred from $\left| 1
\right\rangle $ to $\left| 2 \right\rangle $ matches the transition
from $\left| 2 \right\rangle $ to $\left| 3 \right\rangle $
perfectly.  Further increasing of  ${{\Gamma _s}}$   deteriorates
the   coherent transfer, and eventually the efficiency decreases to
the relaxation-dominated case, i.e., $P = \frac{{{\Gamma
_{12}}}}{{{\Gamma _{12}} + \Gamma}}$ when ${\Gamma _s} \to \infty $.
Thus, excitation transferred from $\left| 1 \right\rangle $ to
$\left| 2 \right\rangle $ mismatches the transition  from $\left| 2
\right\rangle $ to $\left| 3 \right\rangle $. In Ref
\cite{Plenio200810}, only the latter case, i.e.,  ${J^2} > {\Gamma
_{12}}(\Gamma  + {\Gamma _{12}} + {\Gamma _{21}})$,  is discussed
because the relaxation process is not considered, i.e.,
$\Gamma_{12}=\Gamma_{21}=0$.}

The enhancement achieved by the dephasing  can be understood as the
fluctuation-induced-broadening  of energy levels, and this gives
rise to  a conjecture that the fluctuations in the site energies and
couplings may affect the transfer efficiency. We will focus on this
question in the following discussions.

\begin{figure*}[htbp]
\centering
\includegraphics[scale=0.355]{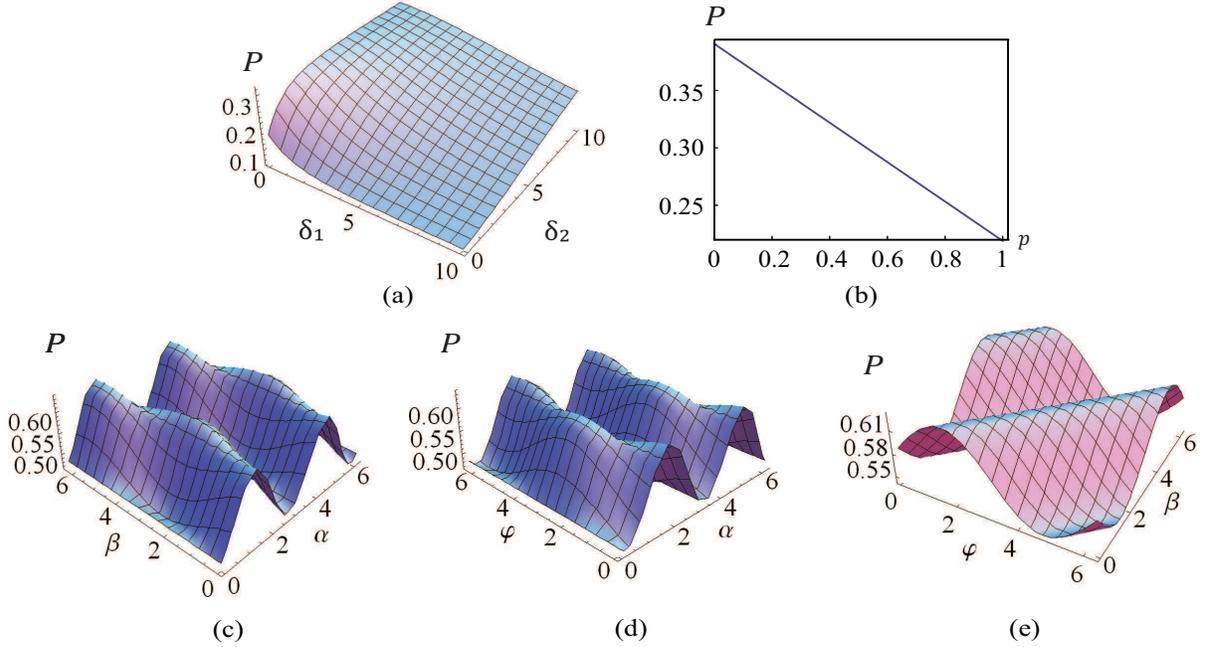}
\caption{(Color online) The transfer efficiency $P$. (a) $P$ vs
${\delta _1}$ and ${\delta _2}$. (b) $P$ vs $p$. (c) $P$ vs $\alpha
$ and $\beta $ ($\varphi  = \frac{\pi }{3}$). (d) $P$ vs $\alpha $ and $\varphi  $ ($\beta  = \frac{\pi }{3}$). (e) $P$ vs
$\beta $ and $\varphi  $ ($\alpha  = \frac{\pi }{3}$). Unless otherwise addressed, the other
parameters chosen  are $\Gamma  = 1$, $\gamma  = 0.1$, $J = 1$,
${\Gamma _s} = 1$, ${\Gamma _{12}} = {\Gamma _{21}} = 1$, $\Delta  =
0$.} \label{phase:}
\end{figure*}
We add two energy fluctuations ${\delta _1}$, ${\delta _2}$, to the
on-site energy gap $\Delta $ and inter-site coupling $J$ separately.
The expression of $P$  Eq. (\ref{solution 2}) can be easily derived,
\begin{eqnarray}
\begin{aligned}
P = \frac{{{\Gamma _s}}}{{(2\Gamma  + {\Gamma _s}) + \frac{{\Gamma ({A^2} + 16{{(\Delta
+ {\delta _1})}^2})(\Gamma  + {\Gamma _s})}}{{(4{{(J + {\delta _2})}^2}A +
 {\Gamma _\kappa }({A^2} + 16{{(\Delta  + {\delta _1})}^2}))}}}},
\label{solution fluctuation}
\end{aligned}
\end{eqnarray}
where $A = 2\gamma  + 2\Gamma  + 2{\Gamma _s} + 2{\Gamma _\kappa }$ and ${\Gamma _{12}} = {\Gamma _{21}} = {\Gamma _\kappa }$ are performed. Obviously, it monotonically increases with ${\delta _2}$(or $J)$,
whereas it  decreases with ${\delta _1}$(or $\Delta )$. This is easy
to be understood. The energy gap $\Delta $ blocks the energy
transfer, while the inter-site coupling $J$ that represents the
overlap of sites $\left| 1 \right\rangle $ and $\left| 2
\right\rangle $ favors the transport. Fig.~\ref{phase:} (a) shows
this result.

\textbf{Now we examine  the effect of phase $\varphi$ on the
transfer efficiency.} $\varphi$ in the coupling is determined by the
spatial distribution of $\left| 1 \right\rangle $ and $\left| 2
\right\rangle $,   this suggests  to consider a superposition of
$\left| 1 \right\rangle $ and $\left| 2 \right\rangle $ as the
initial state. \textbf{For the FMO complex, the spatial and temporal
relaxation of exciton shows that site 1 and 6 were populated
initially  \cite{Adolphs200691} in sunlight. In laboratory, Ref
\cite{Collini2010463} reported  the initial state preparation of the
system in a coherent superposition of the antenna protein¡¯s
electronic vibrational eigenstates with the femtosecond laser pulse
(25-fs duration).} Then it is reasonable to study how the initial
states affect the excitation transfer efficiency. In the following,
we shall shed light on this question and analyze whether the initial
states have effects on the EET efficiency and whether the   phase
$\varphi $ can enhance the transfer. We consider two different
initial states. One is a superposition state  of sites $\left| 1
\right\rangle $ and $\left| 2 \right\rangle $, and the other is a
classical mixture of $\left| 1 \right\rangle $ and $\left| 2
\right\rangle $.

We illustrate numerical  results of the
transfer efficiency as a function of $\alpha $ and $\beta $, which
characterize the pure initial state of the system through $\left|
{\psi (t = 0)} \right\rangle  = \cos \alpha \left| 1 \right\rangle +
\sin \alpha \exp (i\beta )\left| 2 \right\rangle $, where $\alpha $
denotes the population ratio between sites $\left| 1 \right\rangle $
and $\left| 2 \right\rangle $, and $\beta $ characterizes the
relative phase. From Fig.~\ref{phase:} (c), we observe that for
excitation initially excited in a superposition of sites $\left| 1
\right\rangle $ and $\left| 2 \right\rangle $, both the population
ratio and the relative phase affect the energy transfer, and
suitable values of $\alpha $ and $\beta $ facilitate the efficiency.
The efficiency $P$ reaches its maximal value when $\alpha $
approximately equals to $\frac{\pi }{2}$ and $\frac{{3\pi }}{2}$. This tells that occupation
on  site $\left| 2 \right\rangle $ helps the EET efficiency, see Eq.
(\ref{efficiency}). Fig.~\ref{phase:} (d) shows the dependence of
$P$ on $\varphi  $ and $\alpha $, and Fig.~\ref{phase:} (e) shows
the dependence of $P$  on $\varphi  $ and $\beta $. These two
figures indicate that the phase  $\varphi$ does have effect on the
efficiency when the initial state is a superposition of the two
sites. \textbf{This observation can be understood by an analytical solution
\begin{eqnarray}
\begin{aligned}
&P = \\
&\frac{{{\Gamma _s}(8{J^2}  + 2{\Gamma _{12}}A + \Gamma A{{\sin
}^2}\alpha  - 2J\Gamma \sin 2\alpha \sin (\beta  - \varphi
))}}{{(4{J^2}(2\Gamma  + {\Gamma _s}) + A\Gamma (\Gamma  + {\Gamma
_s} + {\Gamma _{21}} - {\Gamma _{12}}))}} \label{solution 4}
\end{aligned}
\end{eqnarray}
obtained from Eq. (\ref{master equation 1}) with $\Delta  = 0$ and
$A = 2\gamma  + 2\Gamma  + {\Gamma _s} + {\Gamma _{12}} + {\Gamma
_{21}}$. The factor ${\sin (\beta  - \varphi )}$ in the numerator
indicates that varying the values of $\varphi$ is equivalent to
varying the values of the relative phase $\beta$. This is actually
equivalent to choose different coherent superposition  of sites
$\left| 1 \right\rangle $ and $\left| 2 \right\rangle $ as the
initial states, for example,
\begin{eqnarray}
\begin{aligned}
\rho (t = 0) = \left( {\begin{array}{*{20}{c}}
{{{\cos }^2}\alpha }&{\sin \alpha \cos \alpha {e^{ - i\beta }}}\\
{\sin \alpha \cos \alpha {e^{i\beta }}}&{{{\sin }^2}\alpha }
\end{array}} \right).
\label{initial state matrix form}
\end{aligned}
\end{eqnarray}
Therefore, when the network is initially in a superposition state,
the effect of $\varphi$ on the EET is obvious. When $\varphi$ takes
a fixed value, the relative phase $\beta $ also affects the
coherence of exciton transfer and the transfer efficiency $P$. But
when the exciton is  initially in site $\left| 1 \right\rangle $,
neither $\varphi$ nor $\beta $ has effect.}

\textbf{In Ref. \cite{Yuen20146}, the concepts of topological phases
is introduced to explore the exciton transport in porphyrin thin
films. This is the first work that addresses the joint effects of
both magnetic field and coherence in molecular exciton transport.
The magnetic field serves to break time-reversal symmetry and
results in lattice fluxes that mimic the Aharonov-Bohm phase
acquired by electrons. The first blueprint for realizing topological
phases of matter in molecular aggregates stimulates us to study the
EET dynamics from  new point of view.}

Another type of  initial states we consider is mixed state with
mixing rate $p$, $p\left| 1 \right\rangle \left\langle 1 \right| +
(1 - p)\left| 2 \right\rangle \left\langle 2 \right|$,  see
Fig.~\ref{phase:} (b). We observe that $p$ works like $\alpha $
denoting the population weight. More population  on site $\left| 2
\right\rangle $ favors the transport, and  $\varphi$ has no effect
on the efficiency. This is because for  sites $\left| 1
\right\rangle $ and $\left| 2 \right\rangle $ being initially in a
classical mixed state, there is no coherence initially, leading to
no influence of $\varphi  $ on the efficiency. Actually, we obtain
an exact analytical expression for $P$ in this case,
\begin{eqnarray}
\begin{aligned}
P = \frac{{{\Gamma _s}}}{{(2\Gamma  + {\Gamma _s})  + \frac{{A\Gamma
\left[ {(2p - 1)\Gamma  + p{\Gamma _s}} \right]}}{{A\left[ {(1 -
p)\Gamma  + {\Gamma _\kappa }} \right] + 4{J^2}}}}}, \label{solution
3}
\end{aligned}
\end{eqnarray}
where $A = 2\gamma  + 2\Gamma  + 2{\Gamma _s}  + 2{\Gamma _\kappa }$
and ${\Gamma _{12}} = {\Gamma _{21}} = {\Gamma _\kappa }$ are
performed. From this expression we easily know that $P$ is
independent of $\varphi  $.

The effect of phase $\varphi$ on the EET dynamics and  efficiency
will be discussed further in section ${\rm V}$, using a
time-convolutionless master equation.
\section{EET dynamics in non-Markovian case}
In section ${\rm III}$, we treat pure dephasing in a way equivalent
to the   Haken-Strobl model, in which the environment is modeled  as
a spatially uncorrelated classical white noise. Now we consider the
EET dynamics using the Lindblad master equation  derived from a
time-convolutionless master equation within the weak-coupling,
Born-Markovian and secular approximations
\cite{Breuer2002,Mohseni2008129,Fassioli20101}. In contrast to the
Haken-Strobl model, the environment is treated in a quantum way,
\begin{eqnarray}
\begin{aligned}
\dot \rho (t) =  - i[{H_S},\rho ] + {L_\Gamma }
(\rho (t)) + {L_s}(\rho (t)) + L(\rho (t)),
\label{master equation 2}
\end{aligned}
\end{eqnarray}
where the first three terms on the right side are the same as those
in  Eq. (\ref{master equation 1}), representing coherent evolution,
dissipation and sink, respectively. $L(\rho (t))$ is the Lindblad
superoperator given by \cite{Rebentrost2009131,Breuer2002}:
\begin{eqnarray}
\begin{aligned}
&L(\rho (t)) =\sum\limits_{m,\omega } {{\gamma}(\omega,t )}\\
&\left[ {{A_m}(\omega )\rho (t)A_m^\dag (\omega )-
\frac{1}{2}\left\{ {A_m^\dag (\omega ){A_m}(\omega ),\rho (t)} \right\}} \right],
\label{Lindblad gamma 2}
\end{aligned}
\end{eqnarray}
where ${{\gamma}(\omega,t )}$ is the non-Markovian decoherence rate under the assumption
of Ohmic spectral density with exponential cutoff
\begin{eqnarray}
\begin{aligned}
J(\omega ) = \frac{{{\lambda}}}{{{\omega _c}}}\omega \exp \left
( { - \frac{\omega }{{{\omega _c}}}} \right),
\label{spectral density}
\end{aligned}
\end{eqnarray}
where ${{\omega _c}}$ is the cutoff frequency and ${{\lambda}}$ is
the strength of the system-bath coupling. Generally, the
non-Markovian decoherence rate is  time-dependent  given by
\begin{eqnarray}
\begin{aligned}
&{\gamma}\left( {\omega,t} \right) =
2\int_0^\infty  {d\tilde \omega J\left( {\tilde \omega } \right)}\\
&\left( {n\left( {\tilde \omega } \right)\frac{{\sin \left[ {\left( {\omega  +
  \tilde \omega } \right)t} \right]}}{{\omega  + \tilde \omega }} +
  \left[ {n\left( {\tilde \omega } \right) + 1} \right]\frac{{\sin \left[ {\left( {\omega  -
  \tilde \omega } \right)t} \right]}}{{\omega  - \tilde \omega }}} \right),
\label{Lindblad gamma 2}
\end{aligned}
\end{eqnarray}
where $n\left( \omega  \right) = {\left[ {\exp \left( {{\omega  \mathord{\left/
 {\vphantom {\omega  {{k_B}T}}} \right.
 \kern-\nulldelimiterspace} {{k_B}T}}} \right) - 1} \right]^{ - 1}}$ is the bosonic
distribution. In the Markovian limit $\left( {t \to \infty }
\right)$, the decoherence rate can be obtained
\begin{eqnarray}
\begin{aligned}
\gamma \left( {\omega ,\infty } \right) =
2\pi J\left( {\left| \omega  \right|} \right)\left| {n\left( { - \omega } \right)} \right|.
\label{Lindblad gamma markovian}
\end{aligned}
\end{eqnarray}
The dephasing rate can be derived from Eq. (\ref{Lindblad gamma 2})
in the limit $\omega \to 0$
\begin{eqnarray}
\begin{aligned}
{\gamma _\varphi  }\left( t \right) =
2\int_0^\infty  {d\tilde \omega J\left( {\tilde \omega } \right)\coth \left
( {\frac{{\tilde \omega }}{{2{k_B}T}}} \right)\frac{{\sin \left( {\tilde \omega t} \right)}}
{{\tilde \omega }}}.
\label{gamma non-markovian dephasing}
\end{aligned}
\end{eqnarray}
In the Markovian limit $\left( {t \to \infty } \right)$, the dephasing rate becomes
\begin{eqnarray}
\begin{aligned}
{\gamma _\varphi }\left( \infty  \right) = \frac{{2\pi {k_B}T\lambda }}{{{\omega _c}}}.
\label{gamma markovian dephasing}
\end{aligned}
\end{eqnarray}
In the later simulation, unless otherwise noticed, we set $\lambda$
= 50 cm$^{ - 1}$, ${\omega _c}$ = 50 cm$^{ - 1}$ as in Ref.
\cite{Rebentrost2009131}, and note that in units of $\hbar $=1, we
have 1 ps$^{ - 1}$ = 5.3 cm$^{ - 1}$.

${{A_m}(\omega )}$ is the jump operators defined by
\begin{eqnarray}
\begin{aligned}
{A_m}(\omega ) = \sum\limits_{{\lambda _k} - {\lambda _l} = \omega }
{c_m^ * (k){c_m}(l)\left| {{\lambda _k}} \right\rangle \left\langle
{{\lambda _l}} \right|}. \label{jump operators}
\end{aligned}
\end{eqnarray}
The sum runs over all possible transitions in the single exciton
manifold. In our model, the   basis $\left| {{\lambda _k}}
\right\rangle  = \sum\limits_{m = 1,2} {{c_m}\left( k \right)}
\left| m \right\rangle \left( {k = 1,2} \right)$ is composed by the
eigenbasis of Hamiltonian (\ref{system H}), ${H_S}\left| {{\lambda
_k}} \right\rangle = {\lambda _k}\left| {{\lambda _k}} \right\rangle
$. Simple algebra yields,
\begin{eqnarray}
\begin{aligned}
\begin{array}{l}
 {\lambda _1} = \frac{1}{2}[{\varepsilon _1} + {\varepsilon _2} +
 \sqrt {{{\left( {{\varepsilon _1} - {\varepsilon _2}} \right)}^2} + 4{J^2}} ], \\
 {\lambda _2} = \frac{1}{2}[{\varepsilon _1} + {\varepsilon _2} -
 \sqrt {{{\left( {{\varepsilon _1} - {\varepsilon _2}} \right)}^2} + 4{J^2}} ]. \\
 \end{array}
\label{eigenenergies}
\end{aligned}
\end{eqnarray}
And the corresponding eigenstates are,
\begin{eqnarray}
\begin{aligned}
\begin{array}{l}
\left| {{\lambda _1}} \right\rangle  =
\sin \theta \left| 1 \right\rangle  + \cos \theta \exp (i\varphi )\left| 2 \right\rangle ,\\
\left| {{\lambda _2}} \right\rangle  =  \cos \theta \left| 1
\right\rangle  - \sin \theta \exp (i\varphi )\left| 2 \right\rangle
.
\end{array}
\label{inverselaplace1}
\end{aligned}
\end{eqnarray}
where the mixing angle is given by
\begin{eqnarray}
\begin{aligned}
{\sin \theta  = \frac{{{\varepsilon _1}  - {\varepsilon _2} + \sqrt {4{J^2} + {{({\varepsilon _1} - {\varepsilon _2})}^2}} }}{{\sqrt {4{J^2} + {{\left( {{\varepsilon _1} - {\varepsilon _2} + \sqrt {4{J^2} + {{({\varepsilon _1} - {\varepsilon _2})}^2}} } \right)}^2}} }}},\\
{\cos \theta  = \frac{{2J}}{{\sqrt {4{J^2} + {{\left( {{\varepsilon _1} - {\varepsilon _2} + \sqrt {4{J^2} + {{({\varepsilon _1} - {\varepsilon _2})}^2}} } \right)}^2}} }}}.
\label{inverselaplace1}
\end{aligned}
\end{eqnarray}
Thus the jump operators for relaxation are given by
\begin{eqnarray}
\begin{aligned}
\begin{array}{l}
{A_{1,2}}\left( {{\omega _{12}}} \right) =  \pm \sin \theta \cos \theta \left| {{\lambda _1}} \right\rangle \langle {\lambda _2}|,\\
{A_{1,2}}\left( {{\omega _{21}}} \right) =  \pm \sin \theta \cos \theta \left| {{\lambda _2}} \right\rangle \langle {\lambda _1}|,
\end{array}
\label{inverselaplace1}
\end{aligned}
\end{eqnarray}
for dephasing, they are
\begin{eqnarray}
\begin{aligned}
\begin{array}{l}
{A_1}\left( 0 \right) = {\sin ^2}\theta \left| {{\lambda _1}} \right\rangle \langle {\lambda _1}| + {\cos ^2}\theta \left| {{\lambda _2}} \right\rangle \langle {\lambda _2}|,\\
{A_2}\left( 0 \right) = {\cos ^2}\theta \left| {{\lambda _1}} \right\rangle \langle {\lambda _1}| + {\sin ^2}\theta \left| {{\lambda _2}} \right\rangle \langle {\lambda _2}|.
\end{array}
\end{aligned}
\end{eqnarray}
In the Hilbert space spanned by the eigenstates of the Hamiltonian,
$L(\rho (t))$ can be written as
\begin{eqnarray}
\begin{aligned}
L\left( {\rho \left( t \right)} \right) =
&\sum\limits_{n = 1,2} {{\gamma _n}({\sigma _{nn}}\rho \left( t \right){\sigma _{nn}} -
\frac{1}{2}\left\{ {{\sigma _{nn}},\rho \left( t \right)} \right\})} \\
&+ {\gamma _{12}}\left( {{\sigma _{11}}\rho \left( t \right){\sigma _{22}} +
{\sigma _{22}}\rho \left( t \right){\sigma _{11}}} \right) \\
&+ {\Gamma _{21}}({\sigma _{21}}\rho \left( t \right){\sigma _{12}} -
\frac{1}{2}\left\{ {{\sigma _{11}},\rho \left( t \right)} \right\}) \\
&+ {\Gamma _{12}}({\sigma _{12}}\rho \left( t \right){\sigma _{21}} -
\frac{1}{2}\left\{ {{\sigma _{22}},\rho \left( t \right)} \right\}),
\label{dephasing in eigenstate representation}
\end{aligned}
\end{eqnarray}
where ${\sigma _{ij}} = \left| {{\lambda _i}} \right\rangle
\left\langle {{\lambda _j}} \right|$. The first two terms represent
the dephasing  and the last two terms describe relaxation, which is
similar to Eq. (\ref{Lindblad kappa}) and Eq. (\ref{Lindblad gamma
1}). The relaxation and dephasing here  is expressed in the basis
composed of the Hamiltonian eigenstate, different from the
 phenomenological treatment  in Eq. (\ref{Lindblad kappa}) and
 Eq. (\ref{Lindblad gamma 1}). The decoherence rates  defined
 in Eq. (\ref{Lindblad gamma 2} - \ref{gamma markovian dephasing})
 take,
\begin{eqnarray}
\begin{aligned}
{\gamma _n} = &({\sin ^4}\theta  + {\cos ^4}\theta ){\gamma _\varphi }\left( t \right),\\
{\gamma _{12}} = &2{\sin ^2}\theta {\cos ^2}\theta {\gamma _\varphi }(t),\\
{\Gamma _{12}} = &2{\sin ^2}\theta {\cos ^2}\theta \gamma \left( {{\omega _{12}},t} \right),\\
{\Gamma _{21}} = &2{\sin ^2}\theta {\cos ^2}\theta \gamma \left( {{\omega _{21}},t} \right),
\end{aligned}
\end{eqnarray}
for non-Markovian dynamics. They take,
\begin{eqnarray}
\begin{aligned}
{\gamma _n} = &({\sin ^4}\theta  + {\cos ^4}\theta ){\gamma _\varphi }\left( \infty  \right),\\
{\gamma _{12}} = &2{\sin ^2}\theta {\cos ^2}\theta {\gamma _\varphi }(\infty ),\\
{\Gamma _{12}} = &2{\sin ^2}\theta {\cos ^2}\theta \gamma \left( {{\omega _{12}},\infty } \right),\\
{\Gamma _{21}} = &2{\sin ^2}\theta {\cos ^2}\theta \gamma \left( {{\omega _{21}},\infty } \right),
\end{aligned}
\end{eqnarray}
for Markovian case, where ${\omega _{kl}} = {\lambda _k} - {\lambda _l}$ is the transition
frequency. The parameters chosen are: $J$ = 87 cm$^{ - 1}$, ${\varepsilon _1} = 0$ and
${\varepsilon _2}$ = 120 cm$^{ - 1}$, equivalent to the Hamiltonian of sites $\left| 1
\right\rangle $ and $\left| 2 \right\rangle $ subsystem in the FMO complex in
Ref. \cite{Ishizaki2009106}.

%Fig1
\begin{figure*}[htbp]
\centering
\includegraphics[scale=0.42]{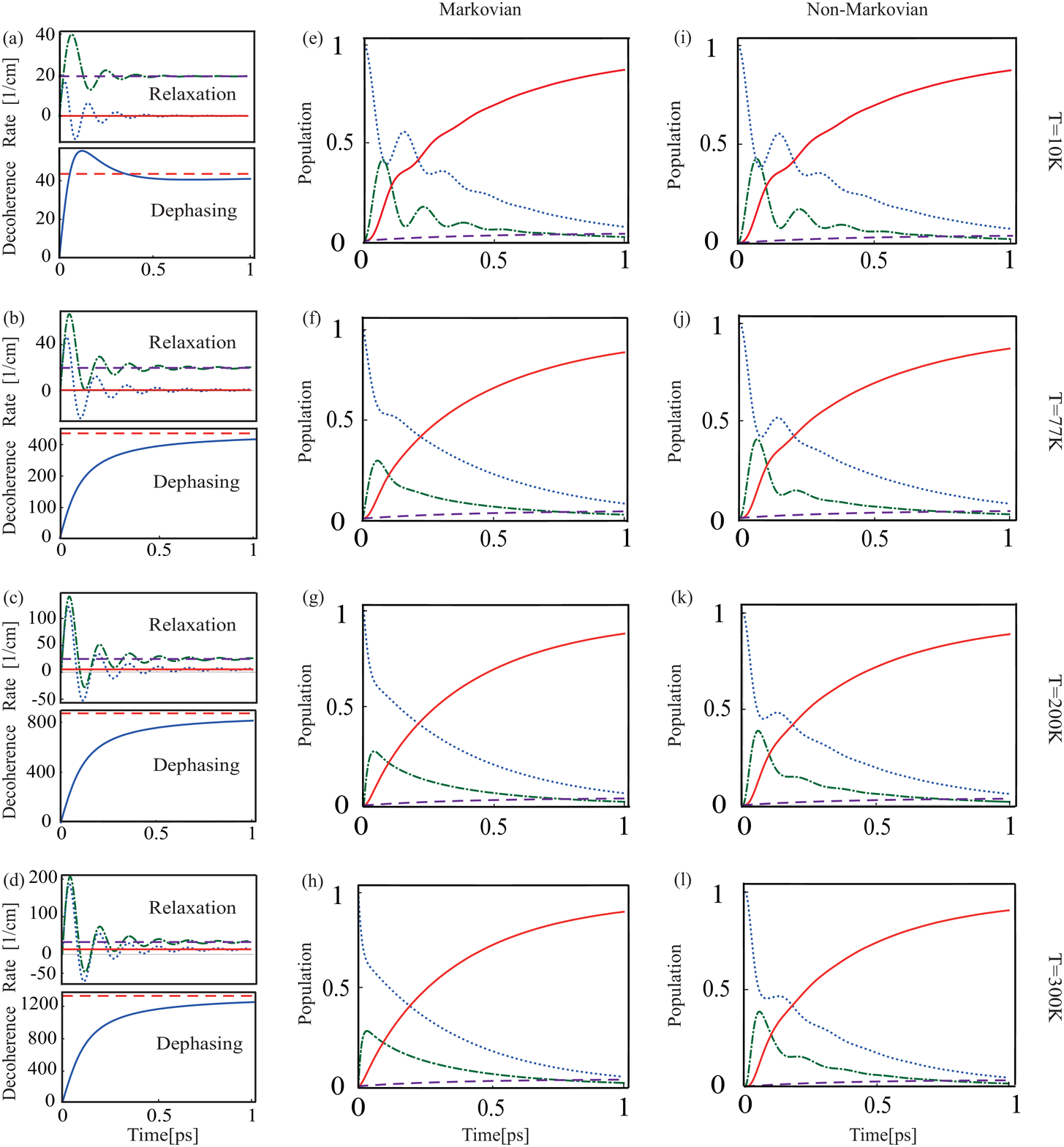}
\caption{(Color online) Time-dependent decoherence rates from Eq.
(\ref{Lindblad gamma 2},\ref{Lindblad gamma markovian},\ref{gamma
non-markovian dephasing},\ref{gamma markovian dephasing}) and time
evolution of population on each site from Eq.(\ref{master equation
2}) for Markovian and non-Markovian dynamics. The left panels
(Fig.~\ref{decoherencerate:} (a) - (d)) show the time evolution of
relaxation (upper panel) and dephasing (lower panel) rates. The
parameters in the  spectral density are ${\omega _c}$ = 50 cm$^{ -
1}$ and $\lambda$ = 50 cm$^{ - 1}$ as  used  in Ref.
\cite{Rebentrost2009131}. The relaxation rates (blue-dotted for
${\omega _{12}}$ and green-dot-dashed for ${\omega _{21}}$)
oscillate, they may take   negative value and converge to values in
the Markovian limit (red-solid for ${\omega _{12}}$, the
purple-dashed line is  for ${\omega _{21}}$). Similarly, the
dephasing rates in the non-Markovian case (blue solid) start with
zero and converge to these in the Markovian limit (red-dashed).
Fig.~\ref{decoherencerate:} (e) - (l) show the population on each
site as a function of time: ${\rho _{00}}$ in purple-dashed, ${\rho
_{11}}$ in blue-dotted, ${\rho _{22}}$ in green-dot-dashed, and
${\rho _{33}}$ in red-solid.   (e) - (h) is plotted for the
Markovian case, and (i) - (l) for the non-Markovian  case.  The
temperatures $T$ and phase $\varphi $ change from figure to figure.
The population on the reaction center ${\rho _{33}}$ with the same
temperatures $T$ is also illustrated  (red-solid for $\varphi   =
0$, orange-dashed for $\varphi = \pi /2$, black dot-dashed for
$\varphi = \pi$). The Hamiltonian is taken to be the same as that in
Ref. \cite{Ishizaki2009106}. Dissipation and trapping rates we chose
are ${\Gamma _1}$ = 0.1 ps$^{ - 1}$, ${\Gamma _2}$ = 0.1 ps$^{ -
1}$, and ${\Gamma _s}$ = 10 ps$^{ - 1}$.} \label{decoherencerate:}
\end{figure*}

%Fig1
\begin{figure*}[htbp]
\centering
\includegraphics[scale=0.42]{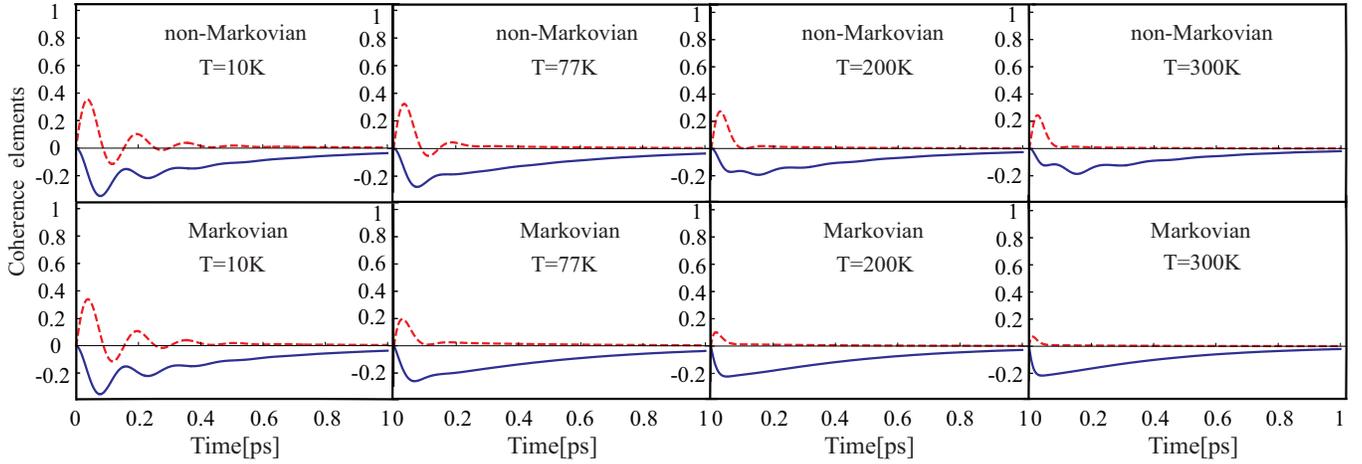}
\caption{(Color online) Time evolution of the  coherence elements
(Re ${\rho _{12}} $ red-dashed, Im ${\rho _{12}}$ blue-solid) from
Eq. (\ref{master equation 2}) for Markovian and non-Markovian
dynamics. The excitonic Hamiltonian is taken to be the form equal to
that in Ref. \cite{Ishizaki2009106}. The
parameters are $\lambda$ = 50 cm$^{ - 1}$, ${\omega _c}$ = 50 cm$^{
- 1}$, phase $\varphi   = 0$, and room temperature $T$ = 300 K.
Dissipation and trapping rates we choose are ${\Gamma _1}$ = 0.1
ps$^{ - 1}$, ${\Gamma _2}$ = 0.1 ps$^{ - 1}$, and ${\Gamma _s}$ = 10
ps$^{ - 1}$. } \label{coherenceelements:}
\end{figure*}

\begin{figure*}[htbp]
\centering
\includegraphics[scale=0.55]{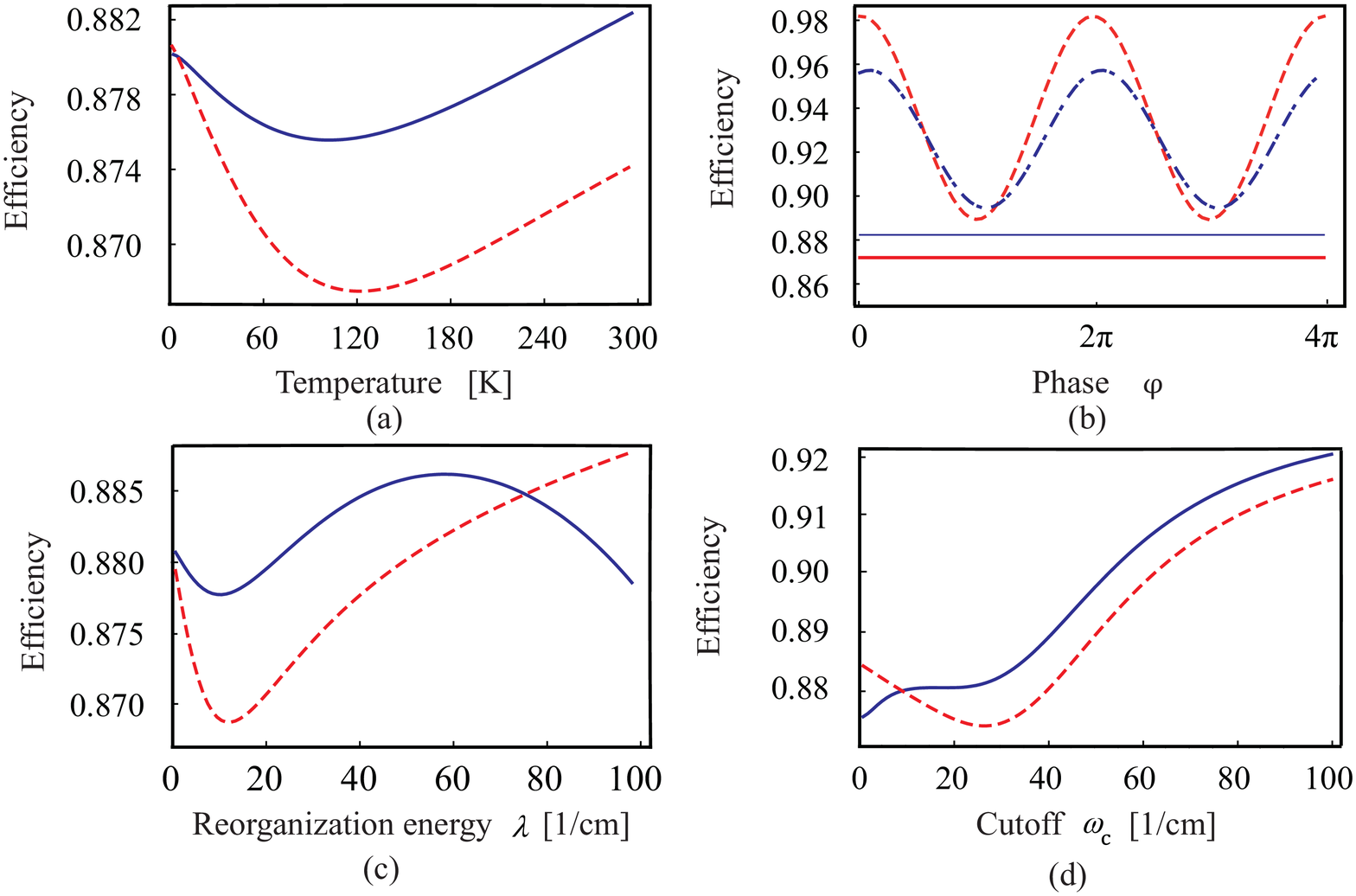}
\caption{(Color online) The transfer efficiency $P$ as a  function
of different temperatures, the phases $\varphi $ and the main
decoherence parameters (reorganization energy $\lambda $, cutoff
${\omega _c}$) for Markovian (red-dashed) and non-Markovian
(blue-soloid) cases. The parameters are ${\omega _c}$ = 30 cm$^{ -
1}$, $\lambda$  = 30 cm$^{ - 1}$, phase $\varphi   = 0$, which is
typical for some natural energy transfer systems like chromophores
in photosynthetic systems, and the temperature is $T$ = 300 K.
Dissipation rate ${\Gamma _{1,2}}$, trapping rate ${\Gamma _s}$,
 and excitonic Hamiltonian takes the same values as
in Fig.~\ref{decoherencerate:}. In (b), thick red-solid for
Markovian and thin blue-solid for non-Markovian with the initial
state ${\rho _{11}}(0) = 1$ , and red-dashed for Markovian and
blue-dot-dashed for non-Markovian with the initial state a
superposition state $\left| {\psi (t = 0)} \right\rangle  =
\frac{1}{2}\left| 1 \right\rangle  + \frac{{\sqrt 3 }}{2}\left| 2
\right\rangle $.} \label{differentparameters:}
\end{figure*}

With these arrangements, we numerically  calculated the time
dependence of the decoherence rates, the time evolution of the
population on the three sites and the coherence element of  the density
matrix in the site basis, i.e., ${\rho _{mn}}\left( t \right) =
\left\langle m \right|\rho \left( t \right)\left| n \right\rangle $.
Markovian and non-Markovian decoherence rates and transfer dynamics
given by Eqs. (\ref{Lindblad gamma 2} - \ref{gamma markovian
dephasing}) are plotted in Fig.~\ref{decoherencerate:} and
Fig.~\ref{coherenceelements:}. The non-Markovian relaxation rates
oscillate, taking positive and negative values, and finally converge
to values in the Markovian limit \cite{Rebentrost2009131}. The dephasing rates in the
non-Markovian case start with zero and similarly, converge to that
in the Markovian limit. We also observe that higher temperature
leads to larger amplitude of oscillation in the relaxation rates.
For the dynamics, the oscillations of population in sites $\left| 1
\right\rangle $ and $\left| 2 \right\rangle $ and  coherence element in the Markovian case
at low temperatures are almost the same with those in the
non-Markovian case. At higher temperatures, however, considerable
differences can be observed from Fig.~\ref{decoherencerate:} and
Fig.~\ref{coherenceelements:}.
Oscillations in the population and coherence element last shortly, whereas in
 the non-Markovian case,
they last long. This can be explained as  follows. For Markovian
case, the dephasing rate is linearly proportional to the temperature
according to Eq. (\ref{gamma markovian dephasing}), whereas at the
same temperature, non-Markovian dephasing rate is much smaller at
short time from Fig.~\ref{decoherencerate:}. In addition, the
Markovian relaxation rate is always positive, while for the
non-Markovian case, the relaxation rate may take positive and
negative. Since the relaxation process represents the diffusion of
excitons through the system, oscillatory relaxation rates of
non-Markovian dynamics lead to beatings of the population and
coherence element, which reflects the memory effect of the
non-Markovian environment.

\textbf{These factors lead to the difference of the  population on
site $\left| 3 \right\rangle $ between Markovian and non-Markovian
cases. Note that we consider the population ${\rho _{33}}$
accumulated on the site 3 for a time of $t$ = 1 ps, i.e., we choose
${\rho _{33}}$ (1 ps) to quantify the transfer efficiency, this is
different from the case in Sec. {\rm III}, where we use ${\rho
_{33}}(\infty )$. Similarly, at low temperatures, the population in
site $\left| 3 \right\rangle $ in Markovian case is almost the same
as in non-Markovian case. At higher temperatures, it exhibits a
slight difference (see Fig.~\ref{decoherencerate:} and
Fig.~\ref{differentparameters:} (a)), which can also be explained by
the different behavior of decoherence rates between Markovian and
non-Markovian cases. Note that  the difference in the efficiency
between Markovian and non-Markovian cases is not so large.
Furthermore, from Fig.~\ref{differentparameters:} (a), we find that
the temperature does not change much the population on  site $\left|
3 \right\rangle $, hence the efficiency is robust against the
temperature. This is consistent with recent experimental results
that the nature photosynthesis are able to transport incident light
energy with nearly $100\% $ efficiency in hot and wet conditions
\cite{Nagy200616,Gilmore2008112,Scholes20113,Grondelle2010463,
Sener2007104,Engel2007446,Lee2007316,Collini2010463,Panitchayangkoon2010107}.}

\textbf{Fig.~\ref{differentparameters:} (b)  shows the  efficiency
as a function of the phase $\varphi  $ for Markovian and
non-Markovian cases. Fig.~\ref{differentparameters:} (b) show that
the phase factor $\varphi  $  has  influence on neither the EET
dynamics nor the efficiency with initial state ${\rho _{11}}(0) =
1$,  this observation holds true for both Markovian and
non-Markovian cases. When we consider a superposition of $\left| 1
\right\rangle $ and $\left| 2 \right\rangle $ as the initial state,
 the effect of $\varphi  $ on the efficiency is clear
from Fig.~\ref{differentparameters:} (b),  and the improvement in
the efficiency  compared with ${\rho _{11}}(0) = 1$ is obvious.
These results are in agreement  with the conclusions in section
${\rm III}$. The effect of the phase factor on the EET depends on
initial states. The difference in the efficiency for Markovian and
non-Markovian cases is very small. }

In Fig. \ref{differentparameters:} (c), we show  the dependence of
transfer efficiency on the reorganization energy $\lambda $ of the
spectral density. From Eq. (\ref{Lindblad gamma 2} - \ref{gamma
markovian dephasing}), we see that relaxation and dephasing rates
are both proportional to $\lambda$. As discussed above, the
dephasing   destroys the   coherence and hinders the energy transfer
when the on-site energy gap $\Delta $, i.e., ${\omega _{12}}$
(${\omega _{21}}$) in this section, is sufficiently small compared
with the dephasing rates, while it improves the energy transfer when
$\Delta $ is suitably large compared with the dephasing rates. But
the relaxation terms describing the exciting and decay processes
among the eigenstates $\left| {{\lambda _1}} \right\rangle $ and
$\left| {{\lambda _2}} \right\rangle $ would only accelerate the
energy transfer. With $\lambda$ getting larger from a very small
value, both the dephasing rates and relaxation rates become larger.
Therefore, the efficiency varies nonmonotonically with $\lambda$
increasing for both Markovian and non-Markovian dynamics.
\textbf{The difference in the EET dynamics for Markovian and
non-Markovian cases results from the different  decoherence rates
when $\lambda $ changes (Fig.~\ref{decoherencerate:} (a) - (d)). The
difference in the transfer efficiency can be even small, see
Fig.~\ref{differentparameters:} (c), this is because  the transfer
efficiency is defined as an accumulation of population on site 3. It
is worth addressing  that the master equation (\ref{master equation
2}) is obtained within the weak-coupling limit. Therefore, for small
values of $\lambda $, the numerical results is reasonable. When
$\lambda $ is large, the master equation (\ref{master equation 2})
is theoretically not applicable, this means it can not be derived as
in textbook, but it is allowed to  use the master equation
phenomenonaly to describe the EET.}

Fig.~\ref{differentparameters:} (d) shows the dependence of $P$ as a
function of ${\omega _c}$. Eq. (\ref{spectral density}) tells that
the spectral density $J(\omega )$ is not a monotonic function of
${\omega _c}$, and the four decoherence rates change differently
with ${\omega _c}$. Similarly, when ${\omega _c}$ changes, the roles
that the relaxation and dephasing play in  the dynamics change for
both Markovian and non-Markovian dynamics, leading to the difference
in  the dependence of $P$ on ${\omega _c}$ between Markovian and
non-Markovian cases, see in Fig.~\ref{differentparameters:} (d).
\textbf{Notice that  a larger cutoff  ${\omega _c}$ means stronger
couplings between the system and environment,  so  the master
equation (\ref{master equation 2}) becomes not applicable from the
respect that it can be derived by the use of week-coupling
approximation. }

The reason why the different surrounding environments  can affect
the EET dynamics but exhibit no influence on the efficiency
(Fig.~\ref{decoherencerate:} - Fig.~\ref{differentparameters:}) lies
in that: when the EET dynamics is in the stage of coherent
oscillation, the exciton is transferred relatively faster. The
surrounding environment and the temperature may influence the
duration of coherent oscillation, but before this stage ends, the
exciton has been transferred to the sink $\left| 3 \right\rangle $
with a large probability, no matter how long this period of coherent
oscillation is. This suggestes that quantum coherence play a
significant role in achieving the remarkable efficient EET
\cite{Amerongen2000,Caruso2009131, Ishizaki2009106}. The explanation
above gives an intuitive picture how the exciton is transferred
through the system which will be further illustrated in terms of the
notion of {\it quantumness} in Sec. {\rm V}.

It is notable that the efficiency of EET for Markovian  and
non-Markovian cases discussed in \cite{Rebentrost2009131} exhibits
considerable difference. It is due to the different measure to
elucidate the EET efficiency. In this paper, we adopt trapping sites
$\left| 3 \right\rangle $ (with trapping rate ${\Gamma _s}$) to
model the sink while \cite{Rebentrost2009131} ignore the trapping
process and utilize another kind of measure given by
\begin{eqnarray}
P = \frac{1}{\tau }\int_0^\tau  {dt\left\langle M \right|} \rho (t)\left| M \right\rangle,
\label{continuity equation}
\end{eqnarray}
where $\tau $ is the total integration time and  $\left| M
\right\rangle $ is a particular site.

\section{Quantumness of energy current in EET}
In Sec.  ${\rm I}{\rm{V}}$, we have shown that quantum coherent
wave-like oscillations of the populations in sites $\left| 1
\right\rangle $ and $\left| 2 \right\rangle $  last  up to 500 fs at
temperature $T$ = 10 K, and become shorter with the
 temperature increasing for both Markovian and non-Markovian dynamics.
In the following,  we will quantify how {\it quantum} the EET
dynamics is and study the influence of environment and spatial
structure of the system on this process. As energy transfer
constitutes the main function of the system, we focus on the
quantumness of the energy transfer current \cite{Nalbach201184}.  As
is widely accepted,  the most classical states of a quantum system
are the pointer states, which are einselected by the decoherence
process. For the case of a measurement of energy current, the
pointer states are the eigenstates of the energy  current operator
\cite{Zurek198124,Zurek200375}. We then can measure the distance of
the density matrix to the convex set of the pointer states
\cite{Giraud201012, Giraud200878, Martin201081}, i.e., classical
mixing the pointer states, and define the  quantumness  as the
minimum distance.

\subsection{Energy currents  in the transport}
First, we derive the energy current operator for our  system
utilizing a well-developed theory by Hardy \cite{Hardy1963132} which
was widely used by a large amount of succeeding literatures
\cite{Deppe199450,Segal2003119,Allen199348,Leitner2009130,Wu200942}.
An energy current operator ${\vec{\rm{j}}}\left( {\rm{x}} \right)$
can be obtained from a continuity equation
\begin{eqnarray}
\dot H\left( {\rm{x}} \right) + \nabla  \cdot {\vec{\rm{ j}}}\left(
{\rm{x}} \right) = 0, \label{continuity equation}
\end{eqnarray}
with $H\left( {\rm{x}} \right)$ the local energy  density. For the
Hamiltonian (\ref{system H}) in tight-binding approximation, $H =
\sum\limits_{m,n}^2 {{h_{mn}}\left| m \right\rangle \left\langle n
\right|} $, we  rewrite the decomposition of the Hamiltonian in
terms of local excitations as
\begin{eqnarray}
\begin{aligned}
&H = {h_1} + {h_2},\\
&{h_m} = \frac{1}{2}\sum\limits_{n = 1}^2
{\left( {{h_{mn}}\left| m \right\rangle \left\langle n \right| + H.c.} \right)},
\label{decomposition of Hamiltonian}
\end{aligned}
\end{eqnarray}
Consider a  one-dimensional chain,  the current can flow only in one
direction, hence $\vec{\rm{ j}}$ has only one component, which will
be denoted  by $j$,  and its value on site $m$  denoted by ${j_m}$.
Taking  as equidistant with lattice constant $a$, we obtain the
discretized form of Eq. (\ref{continuity equation})
\begin{eqnarray}
\begin{aligned}
\frac{\partial }{{\partial t}}\frac{{{h_m}}}{a} = \frac{{j_m^l - j_m^r}}{a},
 \label{one dimensional continuity equations}
\end{aligned}
\end{eqnarray}
with $j_m^{l(r)}$ the energy flux at   positive $x$ direction on the
left (right) of site $m$. The balance of the currents ${s_{1 \to
2}}$ from site $\left| 1 \right\rangle $ to site $\left| 2
\right\rangle $ and ${s_{2 \to 1}}$ from site $\left| 2
\right\rangle $ to site $\left| 1 \right\rangle $ forms current
$j_m^{l(r)}$. With these observations, we have
\begin{eqnarray}
\begin{aligned}
\frac{\partial }{{\partial t}}{h_1} = {s_{2 \to 1}} - {s_{1 \to 2}},\\
\frac{\partial }{{\partial t}}{h_2} = {s_{1 \to 2}} - {s_{2 \to 1}}.
 \label{one dimensional continuity equations 1}
\end{aligned}
\end{eqnarray}
At the same time, we calculate the left-side of Eq. (\ref{one
dimensional continuity equations 1})  by means of Heisenberg's
equation of motion,
\begin{eqnarray}
\begin{aligned}
\frac{\partial }{{\partial t}}{h_m} = i\left[ {H,{h_m}} \right].
\label{Heisenberg's equation}
\end{aligned}
\end{eqnarray}
We have
\begin{eqnarray}
\begin{aligned}
\frac{\partial }{{\partial t}}{h_1} =
\frac{i}{2}\sum\limits_l {\left( {{h_{21}}{h_{1l}}\left| 2
\right\rangle \left\langle l \right| - {h_{12}}{h_{2l}}\left| 1 \right\rangle \left\langle l
\right| - H.c.} \right)},\\
\frac{\partial }{{\partial t}}{h_2} =
\frac{i}{2}\sum\limits_l {\left( {{h_{12}}{h_{2l}}\left| 1
\right\rangle \left\langle l \right| - {h_{21}}{h_{1l}}\left| 2 \right\rangle \left\langle l
\right| - H.c.} \right)}.
\label{one dimensional continuity equations 2}
\end{aligned}
\end{eqnarray}
Comparing with Eq. (\ref{one dimensional continuity equations 1}),
we obtain the  energy current
\begin{eqnarray}
\begin{aligned}
{s_{1 \to 2}} = \frac{i}{2}\sum\limits_l {\left( {{h_{12}}{h_{2l}}\left| 1 \right\rangle
\left\langle l \right| - H.c.} \right)},\\
{s_{2 \to 1}} = \frac{i}{2}\sum\limits_l {\left( {{h_{21}}{h_{1l}}\left| 2 \right\rangle
\left\langle l \right| - H.c.} \right)}.
\label{energy current s}
\end{aligned}
\end{eqnarray}
%Fig2
\begin{figure*}[htbp]
\centering
\includegraphics[scale=0.49]{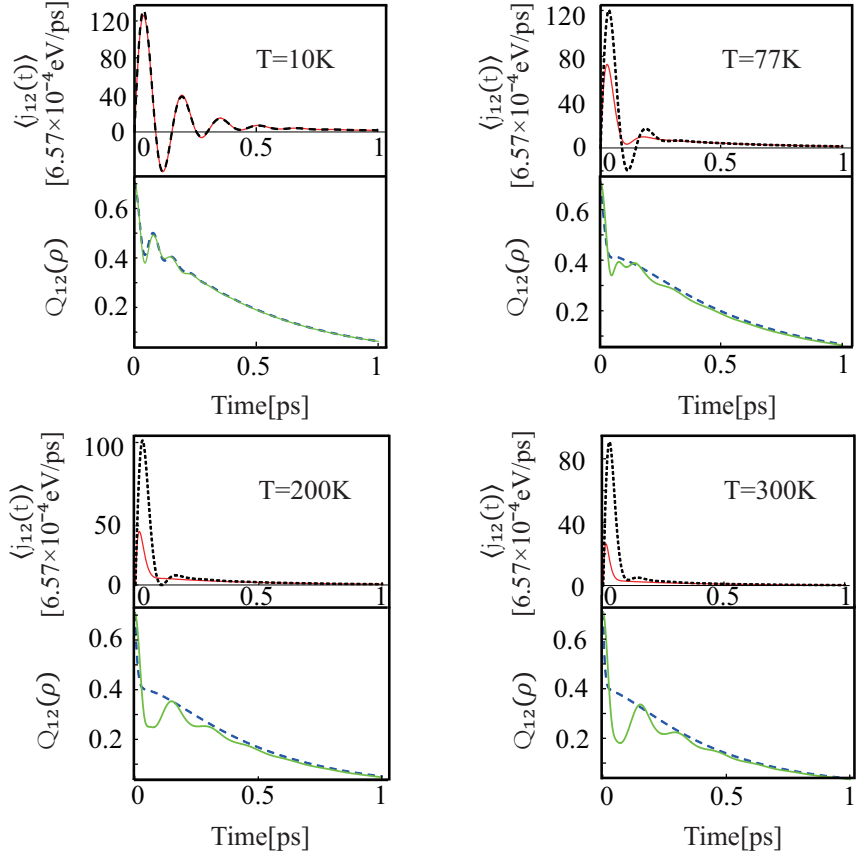}
\caption{(Color online) Energy current  $\left\langle {{j_{12}}(t)}
\right\rangle $ (top) and corresponding quantumness ${Q_{12}}(\rho
)$ (bottom) for the case of Markovian (red-solid for $\left\langle
{{j_{12}}(t)} \right\rangle $ and blue-dashed for ${Q_{12}}(\rho )$)
and non-Markovian (black-dashed for $\left\langle {{j_{12}}(t)}
\right\rangle $ and green-solid for ${Q_{12}}(\rho )$) dynamics with
different values of temperature $T$.
Dissipation rate ${\Gamma _{1,2}}$, trapping rate ${\Gamma _s}$,
spectral density parameters $\lambda $, ${\omega _c}$, in the
excitonic Hamiltonian take the same values as in
Fig.~\ref{decoherencerate:}.} \label{currentquantumness:}
\end{figure*}
The energy current operator is
\begin{eqnarray}
\begin{aligned}
&{j_{12}} = {s_{1 \to 2}} - {s_{2 \to 1}} \\
&=\frac{i}{2}({h_{11}} + {h_{22}})({h_{12}}\left| 1  \right\rangle
\left\langle 2 \right| - {h_{21}}\left| 2 \right\rangle \left\langle
1 \right|). \label{energy current operator j}
\end{aligned}
\end{eqnarray}
Therefore,
\begin{eqnarray}
\begin{aligned}
&\left\langle {{j_{12}}\left( t \right)} \right\rangle  =
tr\left[ {{j_{12}}\rho \left( t \right)} \right] \\&= \frac{i}{2}\left( {{\varepsilon _1} +
{\varepsilon _2}} \right)J\left( {{e^{ - i\varphi  }}{\rho _{21}}\left( t \right) -
{e^{i\varphi  }}{\rho _{12}}\left( t \right)} \right),
\label{energy current j}
\end{aligned}
\end{eqnarray}

where ${{\rho _{12}}\left( t \right)}$  and ${{\rho _{21}}\left( t
\right)}$ are the time evolution of the coherence elements obtained
from Eq. (\ref{master equation 2}). We numerically calculate
$\left\langle {{j_{12}}\left( t \right)} \right\rangle $ at
different temperatures for both Markovian and non-Markovian
dynamics, and present the results in Fig.~\ref{currentquantumness:}.
The result of $\left\langle {{j_{12}}\left( t \right)} \right\rangle
$ is given in unit of $6.57 \times {10^{ - 4}}$ eV/ps. As phase
$\varphi  $ does not affect the EET dynamics  given the initial
state ${\rho _{11}}(0) = 1$, we don't consider the influence of
$\varphi  $.

Lower temperature leads to longer oscillations and larger
oscillation amplitude for both Markovian and non-Markovian cases,
and the amplitude difference between the two cases become more
obvious at larger temperature. This is because the coherence die out
very fast and more  differences between the Markovian and
non-Markovian dynamics emerge at high temperature, which, from Eq.
(\ref{energy current j}), will lead to smaller oscillations and
larger amplitude difference. After the initial time, energy current
becomes small but finite   up to 1 ps, which helps
 in understanding the EET process further based on the
 conclusion obtained in Sec.  ${\rm I}{\rm{V}}$ as follows: energy transfer begins
with quantum coherent population exchange between sites $\left| 1
\right\rangle $, $\left| 2 \right\rangle $ and $\left| 3
\right\rangle $, then small currents gradually redistribute energy
between the three sites and finally the system arrives at an
equilibrium.

\subsection{Quantumness}
The eigenvectors of ${j_{12}}$, i.e., the  pointer  states
\cite{Zurek198124,Zurek200375}, are given by
\begin{eqnarray}
\begin{aligned}
\left| {{\nu ^1}} \right\rangle  = \left( {\begin{array}{*{20}{c}}
   {\frac{{i{e^{ - i\varphi  }}}}{{\sqrt 2 }}}  \\
   {\frac{1}{{\sqrt 2 }}}  \\
\end{array}} \right),
\left| {{\nu ^2}} \right\rangle  = \left( {\begin{array}{*{20}{c}}
   {\frac{{ - i{e^{ - i\varphi  }}}}{{\sqrt 2 }}}  \\
   {\frac{1}{{\sqrt 2 }}}  \\
\end{array}} \right)
\label{eigenvectors of j12}
\end{aligned}
\end{eqnarray}
defining the relevant pure classical states  for the energy
transfer. The minimum distance of ${\rho \left( t \right)}$ to the
convex set of classical states is defined as quantumness of the
current ${j_{12}}$ \cite{Nalbach201184},
\begin{eqnarray}
\begin{aligned}
{Q_{12}}\left( \rho  \right) = {\min _{\left\{ {{p_i}\left| {{p_i} \ge 0,\sum\nolimits_i
{{p_i} = 1} } \right.} \right\}}}\left\| {\rho  - \sum\limits_{l = 1,2} {{p_l}\left|
{{\nu ^l}} \right\rangle \left\langle {{\nu ^l}} \right|} } \right\|,
\label{quantumness}
\end{aligned}
\end{eqnarray}
where $\left\| A \right\|$ is the Hilbert-Schmidt  distance $\left\|
A \right\| = {\left( {trA{A^\dag }} \right)^{{1 \mathord{\left/
{\vphantom {1 2}} \right. \kern-\nulldelimiterspace} 2}}}$.
\textbf{Note that this measure of quantumness is introduced to
characterize  the bipartite  quantum correlations between spatially
separated sites $\left| 1 \right\rangle $ and $\left| 2
\right\rangle $, not the correlation  between the system and
environment \cite{Sarovar20106,Giraud201012,Nalbach201184}. The
effect  of the environment on the quantumness is included  in the
density matrix $\rho$ given by Eq. (\ref{master equation 2}).}
According to this definition, ${Q_{12}}\left( \rho  \right) \ge 0$
where ${Q_{12}}\left( \rho \right) = 0$ if $\rho $ is classical. An
upper bound is given by ${Q_{12}}\left( \rho  \right) \le {Q_{\max
}} \equiv \sqrt {tr{\rho ^2} - {1 \mathord{\left/ {\vphantom {1 d}}
\right. \kern-\nulldelimiterspace} d}} $, with $d$ the dimension of
Hilbert space \cite{Giraud201012}. In our system, ${Q_{\max }}
\simeq 0.707$. A finite value of ${Q_{12}}\left( \rho  \right)$
indicates that there are coherences left in $\rho \left( t \right)$
written in the pointer basis. We numerically calculated the
quantumness ${Q_{12}}\left( \rho  \right)$ for the current
$\left\langle {{j_{12}}\left( t \right)} \right\rangle $ at
different temperatures for both Markovian and non-Markovian
dynamics, and present the results in Fig.~\ref{currentquantumness:},
where ${Q_{12}}\left( \rho \right)$ is in unit of $1$. Either, we
don't consider the influence of $\varphi  $, as we choose ${\rho
_{11}}(0) = 1$ as the initial state.

Similarly, the quantumness ${Q_{12}}\left( \rho  \right)$ shows
oscillations at the beginning of evolution, especially for lower
temperature and non-Markovian case, but it decays fast at higher
temperature. This also can be understood as rapid disappearances of
the coherent  oscillations. It is notable that ${Q_{12}}\left( \rho
\right)$ decreases much slower than the oscillation of the energy
current. From Fig.~\ref{currentquantumness:}, for  $T$ = 10 K,
${Q_{12}}\left( \rho  \right)$ drops from its initial  value
(${Q_{\max }} \simeq 0.707$) to $Q_{12} \simeq 0.3 - 0.4$ within $
\sim 200$ fs, but slowly drops to $Q_{12} \simeq 0.1$ at $ \sim 800$
fs. From Fig.~\ref{decoherencerate:} and
Fig.~\ref{coherenceelements:}, the  population and the off-diagonal
elements of the density matrix does not exhibit coherent features at
$ \sim 800$ fs, while the superposition of the pointer states of the
energy current operator retains substantially coherent, comparing
with the quantumness of the thermal equilibrium state ${\rho _T} =
\exp ( - {H_S}/{k_B}T)$ at the four values of temperature adopted
above, which all go to zero as the irreversible loss of energy to
the sink and environment. That is to say, energy or exciton
transfers through the network to a large extend in the form of a
coherent superposition of pointer states of the energy current
operator. Quantum coherence plays a significant role in achieving
the highly efficient EET. \cite{Amerongen2000,Caruso2009131,
Ishizaki2009106}.

\section{Conclusion}
In conclusion, we have studied the quantum  dynamics of  EET through
a complex quantum network. We show  how the coherent evolution and
the environment-induced decoherence, especially the relaxation, work
together  for the efficient EET. The effect of the phase factor in
the coupling constant on the EET depends on initial states, e.g.,
when the exciton is initially in site $\left| 1 \right\rangle $, the
phase factor $\varphi $ has no effect on the transfer efficiency,
but when the initial state is a superposition of $\left| 1
\right\rangle $ and $\left| 2 \right\rangle $, $\varphi$ really has
effect on the efficiency. Using the Lindblad master equation, we
evaluate the time evolution of population on each site for both
Markovian and non-Markovian cases, we find that the dynamics depends
on the Markovianity of the system but the efficiency is robust
against the parameter variations of the environment. Finally, we
quantified the quantum nature of the EET dynamics by calculating the
energy current in the network and its quantumness, which helps
understand the EET process further. It has been found that the
energy current manifests substantial quantumness in both Markovian
and non-Markovian dynamics. So, to some  extend, we can say the EET
dynamics is coherent despite its coupling to environments.

%%%%%%%%%%%%%%%%%%%%%%%%%%%%%%%%%%%%%%

\section{acknowledgments}
This work is supported by the NSF of China under Grants No 11175032.

\appendix
\section{The derivation of Eq.~(\ref{solution 2})}

In terms of the density matrix elements in the  site basis ${\rho
_{ij}}(t)$, the equation of motion are
\begin{eqnarray}
\begin{aligned}
{{\dot \rho }_{11}}(t) =  & - i(J{e^{ - i\varphi  }}
{\rho _{21}} - J{e^{i\varphi  }}{\rho _{12}}) \\
&-{\Gamma _1}{\rho _{11}} -
{\Gamma _{12}}{\rho _{11}} + {\Gamma _{21}}{\rho _{22}}, \\
 {{\dot \rho }_{22}}(t) = & - i(J{e^{i\varphi  }}{\rho _{12}}
 - J{e^{ - i\varphi  }}{\rho _{21}}) \\
& -{\Gamma _2}{\rho _{22}} + {\Gamma _{12}}{\rho _{11}} - {\Gamma _{21}}
 {\rho _{22}} - {\Gamma _s}{\rho _{22}}, \\
 {{\dot \rho }_{33}}(t) =& {\Gamma _s}{\rho _{22}},\\
 {{\dot \rho }_{00}}(t) =& {\Gamma _1}{\rho _{11}} + {\Gamma _2}{\rho _{22}}, \\
 {{\dot \rho }_{12}}(t) =&  - i[ - 2\Delta {\rho _{12}}
 + J{e^{ - i\varphi  }}({\rho _{22}} -
 {\rho _{11}})] \\
 & -\frac{1}{2}({\Gamma _1} + {\Gamma _2} + {\gamma _1} + {\gamma _2} +
 {\Gamma _{12}} + {\Gamma _{21}} + {\Gamma _s}){\rho _{12}},
 \label{equations}
\end{aligned}
\end{eqnarray}
where $\Delta  = {{({\varepsilon _2} - {\varepsilon _1})} \mathord{\left/
 {\vphantom {{({\varepsilon _2} - {\varepsilon _1})} 2}} \right.
 \kern-\nulldelimiterspace} 2}$, and the initial conditions are
\begin{eqnarray}
\begin{aligned}
{\rho _{11}}(0) = 1,{\rho _{00}}(0) = {\rho _{22}}(0) =
{\rho _{33}}(0) = {\rho _{12}}(0) =
0.
 \label{initialconditions}
\end{aligned}
\end{eqnarray}

By means of Laplace transform, the coupled differential  equations
can be converted into a set of algebraic equations for the Laplace
s-domain variables,
\begin{eqnarray}
\begin{aligned}
s{{\tilde \rho }_{11}} =&  - i(J{e^{ - i\varphi  }}
{{\tilde \rho }_{21}} - J{e^{i\varphi  }}
{{\tilde \rho }_{12}})  \\
&-{\Gamma _1}{{\tilde \rho }_{11}} - {\Gamma _{12}}
{{\tilde \rho }_{11}} + {\Gamma _{21}}{{\tilde \rho }_{22}} + 1, \\
s{{\tilde \rho }_{22}} =&  - i(J{e^{i\varphi  }}{{\tilde \rho }_{12}} - J{e^{ - i\varphi  }}
{{\tilde \rho }_{21}}) \\
& -{\Gamma _2}{{\tilde \rho }_{22}} + {\Gamma _{12}}
{{\tilde \rho }_{11}} - {\Gamma _{21}}{{\tilde \rho }_{22}} - {\Gamma _s}
{{\tilde \rho }_{22}}, \\
s{{\tilde \rho }_{33}} = &{\Gamma _s}{{\tilde \rho }_{22}}, \\
s{{\tilde \rho }_{00}} = &{\Gamma _1}{{\tilde \rho }_{11}} +
{\Gamma _2}{{\tilde \rho }_{22}}, \\
s{{\tilde \rho }_{12}} = & - i[ - 2\Delta {{\tilde \rho }_{12}} + J{e^{ - i\varphi  }}
({{\tilde \rho }_{22}} - {{\tilde \rho }_{11}})]\\
& - \frac{1}{2}({\Gamma _1} + {\Gamma _2} +
{\gamma _1} + {\gamma _2} + {\Gamma _{12}} + {\Gamma _{21}} +
{\Gamma _s}){{\tilde \rho }_{12}}.
 \label{equationslaplace}
\end{aligned}
\end{eqnarray}

From Eq.~(\ref{equationslaplace}), we can easily obtain the
expression of ${{\tilde \rho }_{33}}(s)$, and the relation of the
Laplace  transform for $s$ and $t$ gives
\begin{eqnarray}
\begin{aligned}
P = {\rho _{33}}(\infty ) =
\mathop {\lim }\limits_{s \to 0 } s{{\tilde \rho }_{33}}(s).
 \label{inverselaplace}
\end{aligned}
\end{eqnarray}
Then we obtain the expression of Eq.~(\ref{solution 2}).
Eq.~(\ref{solution 1}), Eq.~(\ref{solution fluctuation}), and
Eq.~(\ref{solution 3}), Eq.~(\ref{solution 4})  can be got in the same way.

\section{Derivative of Eq.~(\ref{solution 1}) with respect to ${\Gamma _s}$}
Now we take the derivative of Eq.~(\ref{solution 1}), and obtain
\begin{eqnarray}
\begin{aligned}
\frac{{dP}}{{d{\Gamma _s}}} = \frac{{B + C{\Gamma _s} + D{\Gamma _s}^2}}{E},
\label{inverselaplace}
\end{aligned}
\end{eqnarray}
where
\begin{eqnarray}
\begin{aligned}
C =& 2\Gamma {\Gamma _{12}}(8{J^2} + (\Gamma  + {\Gamma _{12}} + {\Gamma _{21}}),\\
D =& \Gamma ( - 4{J^2} + {\Gamma _{12}}(\Gamma  + {\Gamma _{12}} + {\Gamma _{21}})),\\
E =& \left[ {(4{J^2} + A{\Gamma _{12}})(2\Gamma  + {\Gamma _s})} \right. \\
&+ {\left. {A\Gamma (\Gamma  + {\Gamma _s} - {\Gamma _{12}} + {\Gamma _{21}})} \right]^2},\\
A =& 2\gamma  + 2\Gamma  + {\Gamma _s} + {\Gamma _{12}} + {\Gamma _{21}},
\label{inverselaplace1}
\end{aligned}
\end{eqnarray}
and $B > 0$ is a very complex expression without ${\Gamma _s}$. From these expressions, the denominator $E > 0$. In the numerator, $C > 0$ and from the quadratic coefficient $D = \Gamma ( - 4{J^2} + {\Gamma _{12}}(\Gamma  + {\Gamma _{12}} + {\Gamma _{21}}))$, when ${J^2} < {\Gamma _{12}}(\Gamma  + {\Gamma _{12}} + {\Gamma _{21}})$, $D > 0$, thus  $P$ is a monotonic function of ${{\Gamma _s}}$. when ${J^2} > {\Gamma _{12}}(\Gamma  + {\Gamma _{12}} + {\Gamma _{21}})$, $D < 0$, then  $P$ is not a monotonic function of ${{\Gamma _s}}$, and it increases first and then decreases.


\begin{references}
\bibitem{Perrin193217} F. Perrin, Ann. Phys. (Paris) \textbf{17}, 283 (1932).

\bibitem{Nagy200616} A. Nagy, V. Prokhorenko, and K. J. Dwayne Miller,
Curr. Opin. Struct. Biol. \textbf{16}, 654 (2006)

\bibitem{Gilmore2008112} J. Gilmore and R. H. McKenzie, J. Phys. Chem.
A \textbf{112}, 2162 (2008).

\bibitem{Scholes20113} G. D. Scholes, G. R. Fleming, A. Olaya-Castro, and
P. V. Grondelie, Nat. Chem. \textbf{3}, 763 (2011); Y. C. Cheng and G. R. Fleming,
Annu. Rev. Phys. Chem. \textbf{60}, 241 (2009).

\bibitem{Grondelle2010463} R. V. Grondelle and V. I. Novoderezhkin, Nature (London)
\textbf{463}, 614 (2010); R. V. Grondelle, J. P. Dekker, T. Gillbro, and V. Sundstrom,
 Biochim. Biophys. Acta \textbf{1187}, 1 (1994).

\bibitem{Sener2007104} M. K. Sener, J. D. Olsen, C. N. Hunter, and K. Schulten,
    Proc. Natl. Acad. Sci. USA \textbf{104}, 15723 (2007).

\bibitem{Engel2007446} G. S. Engel, T. R. Calhoun, E. L. Read, T. K. Ahn,
    T. Manal, Y. C. Cheng, R. E. Blankenship, and G. R. Fleming, Nature (London)
    \textbf{446}, 782 (2007).

\bibitem{Lee2007316} H. Lee, Y. C. Cheng, and G. R. Fleming,
Science \textbf{316}, 1462 (2007).

\bibitem{Collini2010463} E. Collini, C. Y. Wong, K. E. Wilk, P. M. G. Curmi, P. Brumer,
and G. D. Scholes, Nature (London) \textbf{463}, 644 (2010).

\bibitem{Panitchayangkoon2010107}  G. Panitchayangkoon, D. Hayes, K. A. Fransted,
J. R. Caram, E. Harel, J. Wen, R. E. Blankenship, and G. S. Engel,
Proc. Natl. Acad. Sci. USA \textbf{107}, 12766 (2010).

\bibitem{Amerongen2000} H. van Amerongen, L. Valkunas, and R. van Grondelle,
\textit{Photosynthetic Excitons} (World Scientific, Singapore, 2000).

\bibitem{Caruso2009131} F. Caruso, A. W. Chin, A. Datta, S. F. Huelga, and M. B. Plenio, J.
Chem. Phys. \textbf{131}, 105106 (2009).

\bibitem{Ishizaki2009106} A. Ishizaki and G. R. Fleming,
Proc. Natl Acad. Sci. USA \textbf{106}, 17255 (2009).

\bibitem{Plenio200810} M. B. Plenio and S. F. Huelga,
New J. Phys. \textbf{10}, 113019 (2008).

\bibitem{Ai201286} Bao-quan Ai and Shi-Liang Zhu, Phys. Rev. E \textbf{86}, 061917 (2012)

\bibitem{Hoyer201012} S. Hoyer, M. Sarovar, and K. B. Whaley,
New J. Phys. \textbf{12}, 065041 (2010); F. Fassioli and A. Olaya-Castro,
New J. Phys. \textbf{12}, 085006 (2010).

\bibitem{Caruso201285} F. Caruso, S. Montangero, T. Calarco, S. F. Huelga, and
M. B. Plenio, Phys. Rev. A \textbf{85}, 042331 (2012).

\bibitem{Chin201012} A. W. Chin, A. Datta, F. Caruso, S. F. Huelga, and M. B. Plenio,
New J. Phys. \textbf{12}, 065002 (2010).

\bibitem{Cheng200696} Y. C. Cheng and R. J. Silbey,
Phys. Rev. Lett. \textbf{96}, 028103 (2006).

\bibitem{Nazir2009103} A. Nazir, Phys. Rev. Lett. \textbf{103}, 146404 (2009).

\bibitem{Calhoun2009113} T. R. Calhoun, N. S. Ginsberg, G. S. Schlau-Cohen,
Y. C. Cheng, M. Ballottari, R. Bassi, and G. R. Fleming, J. Phys. Chem. B \textbf{113},
16291 (2009).

\bibitem{Rebentrost200911} P. Rebentrost, M. Mohseni, I. Kassal, S. Lloyd, and
A. Aspuru-Guzik, New J. Phys. \textbf{11}, 033003 (2009); P. Rebentrost, M. Mohseni, and
A. Aspuru-Guzik, J. Phys. Chem. B \textbf{113}, 9942 (2009).

\bibitem{Cui201245} B. Cui, X. X. Yi, and C. H. Oh,
J. Phys. B: At. Mol. Opt. Phys. \textbf{45}, 085501 (2012);
B. Cui, X. Y. Zhang, and X. X. Yi, e-print arXiv:1106.4429.

\bibitem{Sarovar20106} M. Sarovar, A. Ishizaki, G. R. Fleming, and K. Birgitta Whaley,
Nature Physics \textbf{6}, 462 (2010).

\bibitem{Ishizaki2009130} A. Ishizaki and G. R. Fleming,
J. Chem. Phys. \textbf{130}, 234111 (2009).

\bibitem{Yang2011132} S. Yang, D. Z. Xu, Z. Song, and C. P. Sun,
J. Chem. Phys. \textbf{132}, 234501 (2011).

\bibitem{Ghosh2011134} P. K. Ghosh, A. Y. Smirnov and F. Nori,
J. Chem. Phys. \textbf{134}, 244103 (2011).

\bibitem{Prior2010105} J. Prior, A. W. Chin, S. F. Huelga, and M. B. Plenio, Phys. Rev. Lett. \textbf{105}, 050404 (2010).

\bibitem{Adolphs200691} J. Adolphs and T. Renger, Biophys. J. \textbf{91}, 2778 (2006).

\bibitem{Cao2009113} J. S. Cao and R. J. Silbey, J. Phys. Chem. A \textbf{113}, 13825 (2009);
A. Kelly and Y. M. Rhee, J. Phys. Chem. Lett. \textbf{2}, 808 (2011).

\bibitem{Faisal20089} A. A. Faisal, L. P. J. Selen and D. M. Wolpert,
Nat. Rev. Neurosci. \textbf{9}, 292 (2008).

\bibitem{Mohseni2008129} M. Mohseni, P. Rebentrost, S. Lloyd, and A. Aspuru-Guzik,
J. Chem. Phys. \textbf{129}, 174106 (2008).

\bibitem{Yi201367} X. X. Yi, X. Y. Zhang, and C. H. Oh,
Eur. Phys. J. D \textbf{67}, 172 (2013).

\bibitem{Haken1973262} H. Haken and G. Strobl , Z. Phys. \textbf{262} 135 (1973).

\bibitem{Rebentrost2009131} P. Rebentrost, R. Chakraborty, and A. Aspuru-Guzik,
J. Chem. Phys. \textbf{131}, 184102 (2009).

\bibitem{Fassioli201012} F. Fassioli and A. Olaya-Castro,
New J. Phys. \textbf{12}, 085006 (2010).

\bibitem{Tan201258} Q.-S. Tan and L.-M. Kuang,
Commun. Theor. Phys. \textbf{58}, 359 (2012).

\bibitem{Liao201082}  J.-Q. Liao, J.-F. Huang, L.-M. Kuang, and C. P. Sun,
Phys. Rev. A \textbf{82}, 052109 (2010).

\bibitem{Briegel0806} H. J. Briegel and S. Popescu, e-print arXiv:0806.4552.

\bibitem{Cai2010104} J. Cai, G. G. Guerreschi, and H. J. Briegel,
Phys. Rev. Lett. \textbf{104}, 220502 (2010).

\bibitem{Thorwart2009478} M. Thorwart, J. Eckel, J. H. Reina, P. Nalbach, and S. Weiss,
Chem. Phys. Lett. \textbf{478}, 234 (2009).

\bibitem{Caruso201081} F. Caruso, A. W. Chin, A. Datta, S. F. Huelga, and M. B. Plenio,
Phys. Rev. A \textbf{81}, 062346 (2010).

\bibitem{Olaya200878} A. Olaya-Castro , C. F. Lee , F. F. Olsen, and N. F. Johnson,
Phys. Rev. B \textbf{78}, 085115 (2008).

\bibitem{Giraud201012} O. Giraud, P. Braun, and D. Braun,
New J. Phys. \textbf{12}, 063005 (2010).

\bibitem{Zurek198124} W. H. Zurek, Phys. Rev. D \textbf{24}, 1516 (1981).

\bibitem{Zurek200375} W. H. Zurek, Rev. Mod. Phys. \textbf{75}, 715 (2003).

\bibitem{Yuen20146} J. Yuen-Zhou, S. K. Saikin, N. Yao, and A. Aspuru-Guzik, e-print arXiv:1406.1472.

\bibitem{Fassioli20101} F. Fassioli, A. Nazir, and A. Olaya-Castro, J. Phys. Chem. Lett.
 \textbf{1}, 2139 (2010).

\bibitem{Breuer2002} H.-P. Breuer and F. Petruccione,
\textit{The Theory of Open Quantum Systems} (Oxford University Press, Oxford, 2007).

\bibitem{Nalbach201184} P. Nalbach, D. Braun, and M. Thorwart, Phys. Rev. E \textbf{84} , 041926 (2011).

\bibitem{Giraud200878} O. Giraud, P. Braun, and D. Braun,
Phys. Rev. A \textbf{78}, 042112 (2008).

\bibitem{Martin201081} J. Martin, O. Giraud, P. A. Braun, D. Braun, and T. Bastin,
Phys. Rev. A \textbf{81}, 062347 (2010).

\bibitem{Hardy1963132} R. J. Hardy, Phys. Rev. \textbf{132}, 168 (1963).

\bibitem{Deppe199450} J. Deppe and J. L. Feldman, Phys. Rev. B \textbf{50}, 6479 (1994).

\bibitem{Segal2003119} D. Segal, A. Nitzan, and P. Hanggi,
J. Chem. Phys. \textbf{119}, 6840 (2003).

\bibitem{Allen199348} P. B. Allen and J. L. Feldman, Phys. Rev. B \textbf{48}, 12581 (1993).

\bibitem{Leitner2009130} D. M. Leitner, J. Chem. Phys. \textbf{130}, 195101 (2009).

\bibitem{Wu200942} L. Wu and D. Segal, J. Phys. A \textbf{42} , 025302 (2009).

\end{references}
\end{document}